\documentclass[aps,prd,preprint,groupedaddress, amsfonts,amssymb,amsmath, floatfix]{revtex4}

\usepackage{bm}
\usepackage{graphicx}

\begin{document}

\title{Dark Energy and Structure Formation: Connecting the Galactic
  and Cosmological Length Scales}  

\author{A.~D.~Speliotopoulos}

\email{achilles@cal.berkeley.edu}

\affiliation{
Department of Mathematics,
Golden Gate University,
San Francisco, CA 94105
}

\affiliation{
Department of Physics,
Ohlone College,
Fremont, CA 94539-0390
}

\date{December 4, 2007}

\begin{abstract}

On the cosmological length scale, recent measurements by WMAP have
validated $\Lambda$CDM to a precision not see before in
cosmology. Such is not the case on galactic length scales, however,
where the `cuspy-core' problem has demonstrated our lack of
understanding of structure formation. Here, we propose a solution to
the 'cuspy-core' problem based on the observation that with the
discovery of Dark Energy, $\Lambda_{DE}$, there is now a universal
length scale, $\lambda_{DE}=c/(\Lambda_{DE} G)^{1/2}$, associated with
the universe. This length scale allows for an extension of the
geodesic equations of motion that affects only the motion of massive
test particles; the motion of massless test particles are not
affected, and such phenomenon as gravitational lensing remain
unchanged. An evolution equation for the density profile is derived,
and an effective free energy density functional for it is
constructed. We conjecture that the pseudoisothermal profile is
preferred over the cusp-like profile because it has a lower effective
free energy. A cosmological check of the theory is made using the
observed rotational velocities and core sizes of 1393 spiral
galaxies. We calculate $\sigma_8$ to be $0.68_{\pm0.11}$; this is
within experimental error of the WMAP value
$0.761_{-0.048}^{+0.049}$. We then calculate $R_{200}=270_{\pm 130}$
kpc, which is in agreement with observations. We estimate the
fractional density of matter that cannot be determined through gravity
to be $0.197_{\pm 0.017}$; this is nearly equal to the WMAP value for the
fractional density of nonbaryonic matter
$0.196^{+0.025}_{-0.026}$. The fractional density of matter that can
be determined through gravity is then calculated to be
$0.041_{-0.031}^{+0.030}$; this is nearly equal to
$\Omega_B=0.0416_{-0.0039}^{+0.0038}$. 
 
\end{abstract}


\maketitle


\section{Introduction}

The recent discovery of Dark Energy (see \cite{Ries1998, Perl1999}
and references therein) has broadened our knowledge of the universe, and
has demonstrated once again that it can hold surprises for
us. The discovery has, most assuredly, also brought into 
sharp relief the degree of our understanding of it. Only a small
fraction of the mass-energy density of the universe is made up of 
matter that we have characterized; the rest consists of Dark Matter
and Dark Energy, and the precise properties of either one is not
known. They are nevertheless needed to explain what is seen on an 
extremely wide range of length scales. On the galactic ($\sim 100$ kpc
parsec), galactic ($\sim$ ~10 Mpc), and  supercluster
($\sim$ 100 Mpc) scales, Dark Matter has been used to explain phenomena
ranging from the formation of galaxies and their rotation curves, to
the dynamics galaxies and the formation galactic clusters and
superclusters. On the cosmological length scale, Dark 
Matter and Dark Energy is needed to explain the observed evolution of
the universe from the Big Bang to the present, and will determine its
fate in the future. 

Observations thus tell us that over a vast range of length scales the
dynamics and evolution of the observed universe is determined not by
normal matter, but by Dark Matter and now Dark Energy (see \cite{Cahi}
for a quantum-cosmology approach that does not require Dark Matter or
Dark Energy). Yet while the need and invocation of Dark Matter is
ubiquitous on a wide range of length scales, our understanding of how
matter determines dynamics at the galactic length scale is
lacking. Recent measurements by WMAP \cite{WMAP} have validated the
$\Lambda$CDM model of cosmology to an precision not seen before in
cosmology. The situation on the galactic scale is not nearly as
settled, however. Here, the cuspy-core problem for the density
profile of matter in galaxies (\cite{JNav, Krav, Moore}, and
\cite{PeeblesRev, Silk} for reviews) has demonstrated our lack of
understanding of the formation of galactic rotation curves.

Current understanding of the structure formation is based on the
work of Peebles \cite{Peebles1984}, where the seeds of galaxies are
due to local fluctuations in the density of matter that grow as the
universe expands. Analytical solutions of this model 
have been done \cite{Gunn, Fill, Hoff1985, Hoff1988} for a number of
special cases, and have resulted in density profiles that are
sensitive to initial conditions, and have a power-law dependence whose
exponents vary over a range of values. More recently, numerical 
simulations \cite{JNav, Krav, Moore} of galaxy formation have been
done, and have consistently resulted in density profiles with a
cusp-like structure  
\begin{equation}
\rho_{\hbox{\scriptsize Simulation}}
=\frac{\rho_i}{(r/R_S)^\gamma(1+(r/R_S)^\alpha)^{(\beta-\gamma)/\alpha}},  
\label{cusp}
\end{equation}  
instead of the expected pseudoisothermal density profile. Here,
$\rho_i$ is a density parameter, $R_s$ is related to the radius of  
the galactic core, and the exponents $(\alpha, \beta, \gamma)$ take
a range of values from $(1.5, 2, 1.5)$ for Moore
et.~al.~\cite{Moore} to $(1.0, 3.0, 1.0)$ for Navarro, Frenk and Wilson
(NFW) \cite{JNav} (see \cite{Silk} for review). This lead Moore
to state in \cite{Moore} that cold dark matter fails to reproduce the galactic 
rotation curves for dark-matter-dominated galaxies, one of
the key reasons that dark matter was proposed in the first place. Soon
afterwards, de Blok and coworkers \cite{Blok-1, Blok-2, McGa} explicitly
demonstrated that the NFW density profile does not fit the density
profile observed for Low Surface Brightness (LSB) galaxies (see also
\cite{Gent-2007} for a recent analysis of cusp structure). Rather, the  
traditional pseudoisothermal profile, with $(\alpha, \beta, \gamma)=
(2,2,0)$, is the better fit. This demonstration is especially
compelling as it is believed that dark matter dominates dynamics
in LSB galaxies.  

There have been a number of attempts to solving the cuspy-core problem
within $\Lambda$CDM \cite{Bode, Dave, Somm, Sper2000}, and they have
had varying degrees of success (see \cite{PeeblesRev} for a review).
This problem does not exist in Milgrom's Modified Newtonian Dynamics
(MOND) \cite{Mil-1, Mil-2, Mil-3} (see \cite{Sand} for a
review)\textemdash a theory where Dark Matter is not needed\textemdash
but there are a number of theoretical and observational problems that
MOND must overcome (see \cite{Sell} for arguments in support of
MOND, however).  

Our approach to solving the cuspy-core problem, and to structure
formation in general, is much more drastic; therefore, its reach is
correspondingly broader. It is based on the observation
that with the discovery of Dark Energy, $\Lambda_{DE}$, we now have a
\text{universal} length scale, $\lambda_{DE} = c/(\Lambda_{DE}G)^{1/2}$,
on hand \footnote{It is possible to also construct the length scale
  $(\hbar c/\Lambda_{DE})^{1/4}\approx 85$ $\mu$ m. Experiments have been 
  shown that this scale does not affect the Newtonian potential.
  \cite{Adel2007}}. The geodesic equations of motion (GEOM)\textemdash 
  and thus the geodesic action\textemdash is no longer unique, and
  extensions of it through the introduction of functions of $Rc^2/\Lambda_{DE}G$
  can be made. While there have been attempts at proving that the
  GEOM are the unique consequence of the Einstein field
  equations \cite{Einstein, Gero1972, Gero2003}, such proofs assume
  that the background metric remains fixed under the passage of the
  test particle.  As our extension of the depends explicitly on the
  energy-momentum tensor of matter\textemdash which includes the
  motion of the test particle\textemdash these proofs do not preclude
  our extension of the GEOM. 

In form, our extension of the GEOM preserves the equivalence principal, and
through the choice of the function of $Rc^2/\Lambda_{DE}G$, we can
insure that their effects are not measurable on terrestrial
scales. Physically, that this choice is possible 
is because $\Lambda_{DE} = (7.21^{ +0.83}_{-0.84}) \hbox{ x }
10^{-30}$ $g/cm^3$ (from \cite{WMAP}) is so much smaller than any
density of matter either presently achievable experimentally, or
present in regions of space accessible to 
experiment. Correspondingly, $\lambda_{DE} = 14020^{790}_{810}$ Mpc is
much longer than the scale of any experiment that has been used to
test general relativity. In fact, the issue is to make the theory
relevant at \textit{galactic} scales of a few kpc, and by doing so we
arrive at an estimate for the exponent $\alpha_\Lambda$. This exponent
is the only parameter in the theory, and it determines the power-law
behavior of our extension of the GEOM. We also find that while
affecting the motion of massive test particles, our extension does not
affect the motion of massless test particles; photons still travel
along null geodesics, and gravitational lensing and the deflection of
light are left unchanged. 
    
Applied to galaxy formation, the extended GEOM reduces to a
nonlinear evolution equation for the density profile of a model galaxy
in the nonrelativistic, linear gravity limit. This evolution
equation minimizes a functional of the density, which is interpreted
as an effective free energy functional for the system. Solutions to 
this equation is found using various velocity curves for galaxies
as driving terms, and these solutions are then used to calculate the
free energy associated with various profiles. We conjecture that
like Landau-Ginzberg phenomenological theories in condensed matter
physics, the system prefers to be in a state that minimizes this free
energy. Showing that the pseudoisothermal profile is preferred over
cusp-like profiles reduces to showing that it has the lower free
energy.   

In our model of a galaxy, the Hubble length scale $\lambda_H =
c/H$ (where $H = h H_0$, is the Hubble constant, $h =
0.732^{+0.031}_{- 0.032}$ and $H_0 = 100$ $km/s/$Mpc \cite{WMAP})
naturally appears, \textit{even though a cosmological model is not
  mentioned either in its construction or in its analysis}. What
happens at galactic length scales are naturally tied 
to what happens  at cosmological length scales with our approach; the
combination of the Dark Energy length scale and the nonlinear aspects
of the extended GEOM link the two. This linkage allows us to
extrapolate from the statistical properties of the observed 
universe the properties of a representative galaxy. These properties
are then used to provide a cosmological check of the theory. 

As with the Peebles model, the total density of matter
for our model galaxy can be written as a sum of a background
  density $\rho_{\hbox{\scriptsize asymp}}(\mathbf{x})$ and a linear
  perturbation $\rho_{II}^{1}(\mathbf{x})$. As usual,
  $\rho_{\hbox{\scriptsize asymp}}(\mathbf{x})$ does not contribute to
  the motion of stars within the galaxy, while
  $\rho_{II}^{1}(\mathbf{x})$ does; in the absence of forces other
  than gravity, observations of the dynamics of
  the stars will determine only $\rho_{II}^{1}(\mathbf{x})$.
  But unlike the Peebles model, $\rho_{\hbox{\scriptsize{asymp}}}$ is
  not a constant. With it, we are able to estimate
  $\Omega_{\hbox{\scriptsize{asymp}}}$, the fractional density of
  matter that \textit{cannot} be determined through gravity, to be 
  $0.197_{\pm0.017}$; this is nearly equal
  to the measured value of the fractional density of nonbaryonic
  (dark) matter in the universe $\Omega_m-\Omega_{B} =
  0.196^{+0.025}_{-0.026}$ measured by WMAP
  \cite{WMAP}. Correspondingly, we estimate
  $\Omega_{\hbox{\scriptsize{Dyn}}}$, the fractional density of matter
  in the universe that \textit{can} be determined through gravity, to
  be $0.041^{+0.030}_{- 0.031}$; this is nearly equal to the
  value of $0.0416^{+0.0038}_{- 0.0039}$ for $\Omega_B$ measured by WMAP  
 \cite{WMAP}. We have also calculated $\sigma_8$, the rms
 fluctuation in the fractional density of matter within a distance of
 $8 h^{-1}$ Mpc, as a direct check of our model. Using the average
 rotational velocity and core sizes of 1393 galaxies obtained through
 four different set of observations \cite{Blok-1, Blok-2, 
   McGa, Rubin1980, Rubin1982, Burs, Rubin1985, Cour, Math} that span
 25 years, we obtain the value of $0.68_{\pm 0.11}$ for $\sigma_8$;
 this value is within experimental error of the measured value of
 $0.761^{+0.049}_{-0.048}$ by WMAP. Finally, we have calculated $R_{200}$,
 the radius of the galaxy at which the density equals 200 
times that of the critical density, to be $270_{\pm 130}$ kpc; this
value also agrees well with observations. 

Interestingly, $\Omega_{\hbox{\scriptsize asymp}}/\Omega_\lambda$
depends only on the dimensionality and symmetry of spacetime, and the
exponent $\alpha_\Lambda$. This suggests that there is an underlying
coupling between Dark Energy and matter in the theory. Such a
coupling has been dismissed before, primarily because it is believed
that the coupling would result in a ``fifth force'' that would already
have been observed \cite{PeeblesRev}. The results of this paper suggest
that with a suitable choice of this coupling, its effects will
not be currently measurable.

A summary of the results here have appeared elsewhere \cite{ADS}. Here,
we provide the details of the theory and the calculations.
 
\section{Extending the Geodesic Lagrangian}

While there is no consensus as to the nature of Dark Energy\textemdash
whether it is due to the cosmological constant $\Lambda_{DE}$ or to
quintessence \cite{Peebles1988, Stei1999, Stei2000}\textemdash
modifications to Einstein's equations 
\begin{equation}
  R_{\mu\nu} - \frac{1}{2}g_{\mu\nu}R +
  \frac{\Lambda_{DE}G}{c^2} g_{\mu\nu}= - \frac{8\pi G}{c^4} T_{\mu\nu},
\label{EinsteinEquation}
\end{equation}
(where $T_{\mu\nu}$ is the energy-momentum tensor for matter,
$R_{\mu\nu}$ is the Ricci tensor, $R$ is the Ricci scalar, Greek
indices run from $0$ to $3$, and the signature of $g_{\mu\nu}$ is
$(1,-1,-1,-1)$) to include the cosmological constant are well known,
and minimal. We only require that $\Lambda_{DE}$ changes so
slowly that it can be considered a constant. We note also that the action
  for gravity+matter is a linear combination of the Hilbert
  action with a cosmological constant term, and the action for
  matter. Any change to the equations of motion for test particles can
  thus be accounted for in $T_{\mu\nu}$, and will not change the
  \textit{form} of Eq.~$(\ref{EinsteinEquation})$. 

With the geodesic Lagrangian
\begin{equation}
\mathcal{L}_0 \equiv mc \left(g_{\mu\nu}\frac{d x^\mu}{dt}\frac{d
  x^\nu}{dt}\right)^{1/2},
\label{geoL}
\end{equation}
and the GEOM
\begin{equation}
v^\nu\nabla_\nu v^\mu \equiv \frac{D v^\mu}{\partial t} = 0,
\label{geoEOM}
\end{equation}
(where $v^\nu = \dot{x}^\nu$ is the four-velocity of the test particle),
it is straightforward to see that in the absence of Dark Energy,
Eq.~$(\ref{geoEOM})$ is the most general form that a second-order evolution
equation for a test particle can take that still obeys the equivalence
principle. Any extension of $\mathcal{L}_0$ requires a dimensionless,
scalar function of some a fundamental property of the spacetime 
  folded in with some physical property of matter. In our
  homogeneous and isotropic universe, there are few opportunities to do
  this. A fundamental vector certainly does not exist in the spacetime,
  and while there is a scalar (the Ricci scalar $R$) and three
  tensors ($g_{\mu\nu}$, the Riemann tensor $R_{\mu\nu,\alpha \beta}$,
  and the Ricci tensor $R_{\mu\nu}$), $R_{\mu\nu,\alpha\beta}$  has
  units of inverse length squared. It is possible to 
  construct a dimensionless scalar  $m^2 G^2 R/c^4$ for the test
  particle, but augmenting 
  $\mathcal{L}_0$ using a function of this scalar would introduce
  additional forces that will depend on the mass of the test
  particle, and thus violate the uniqueness of free fall
  principle. It is also possible to construct the scalar
  $g_{\mu\nu}v^\mu v^\nu/c^2$, but because of the mass-shell condition
  $v_\mu v^\mu=c^2$, any such extension of $\mathcal{L}_0$ will not
  change the GEOM. Scalars may also be constructed from
  $R_{\mu\nu}$ and powers of $R_{\mu\nu,{\alpha_\Lambda}\beta}$ by
  contracting them with the appropriate number of $v^\mu/c$'s, but these
  scalars will once again have dimension of inverse length raised to
  some power, and, as with the Ricci scalar, once again the
  rest mass $m$ is needed to construct a dimensionless quantity. 

For a nonzero Dark Energy, the situation changes dramatically. With a
universal length scale $\lambda_{DE}$, it is now possible to construct
from the Riemann tensor and its contractions dimensionless scalars of
the form, 
\begin{equation}
\frac{c^2R}{\Lambda_{DE}G},\quad \frac{R_{\mu\nu}v^\mu
  v^\nu}{\Lambda_{DE}G}, \quad \frac{c^2v^\mu
  v^\nu}{\left(\Lambda_{DE}G\right)^2} R_{\mu{\alpha}, \beta\gamma}
  R_\nu^{\>\>\>{\alpha}, \beta\gamma}, \quad
  \frac{v^\mu v^\nu
  v_\gamma v_\delta}{\left(\Lambda_{DE}G\right)^2}R_{\mu{\alpha}, \nu\beta}
  R^{\gamma {\alpha}, \delta \beta}, \qquad \dots\>\>\> .
\label{possibles}
\end{equation} 
While extensions to $\mathcal{L}_0$ can be constructed with any of
these terms, we are primarily interested in the nonrelativistic, 
linearized gravity limit. In this limit, the first two terms are
equivalent to one another, while the other terms are smaller than the
first two by powers of $R$, and can be neglected. We therefore
focus solely on the first term, and arrive at the extension
\begin{equation}
\mathcal{L}_{\hbox{\scriptsize{Ext}}} \equiv
mc\Big[1+\mathfrak{D}\left(Rc^2/ \Lambda_{DE}G\right)\Big]^{1/2}
\left(g_{\mu\nu}\frac{d x^\mu}{dt}\frac{d x^\nu}{dt}\right)^{1/2} \equiv
\mathfrak{R}[Rc^2/\Lambda_{DE}G] \mathcal{L}_0,
\label{extendL}
\end{equation} 
for $\mathcal{L}_0$ with the added constraint that $v^2=c^2$ for massive
test particles, and $v^2=0$ for massless test particles. 

If $\mathfrak{D}(x)$ is the constant function, then
$\mathcal{L}_{\hbox{\scriptsize{Ext}}}$ differs from $\mathcal{L}_0$
by an overall constant that can be absorbed through a
reparametization of time. Only non-constant 
$\mathfrak{D}(x)$ are relevant; it is how \textit{fast}
$\mathfrak{D}(x)$ changes that will determine its effect on the
equations of motion. Indeed, in extending $\mathcal{L}_0$ we have
essentially replaced the constant rest 
mass $m$ of the test particle with a curvature-dependent rest mass
$m\mathfrak{R}\left[Rc^2/ \Lambda_{DE}G\right]$. All 
\textit{dynamical} effects of this extension can therefore be
interpreted as the rest energy gained or lost by the test particle due to
the local curvature of the spacetime. The scale of these
effects is of the order of $mc^2/L$, where $L$ is some relevant length
scale of the dynamics, and thus the additional forces from  
$\mathcal{L}_{\hbox{\scriptsize{Ext}}}$ are potentially very \textit{large}. For
these effects \textit{not} to have already been seen is for
$\mathfrak{D}(Rc^2/\Lambda_{DE}G)$ to change very slowly at
current experimental limits.

\subsection{The Extended GEOM for Massive Test Particles}

For massive particles, the extended GEOM from
$\mathcal{L}_{\hbox{\scriptsize{Ext}}}$ is
\begin{equation}
\frac{D^2 x^\mu}{\partial t^2} = c^2 \left(g^{\mu\nu} - \frac{v^\mu
  v^\nu}{c^2}\right)
  \nabla_\nu\log\mathfrak{R}[Rc^2/\Lambda_{DE}G],
\label{genEOM}
\end{equation}
where we have explicitly used $v^2=c^2$. It has a canonical momentum
with a
\begin{equation}
p^2 = p_\mu p^\mu =m^2c^2 \bigg[
1+\mathfrak{D}(Rc^2/\Lambda_{DE}G)\bigg],
\label{extendMass}
\end{equation}
and the interpretation of $m\mathfrak{R}[Rc^2/\Lambda_{DE} G]$ as an
effective rest mass can readily be seen.

The dynamical implications of the new terms in Eq.~$(\ref{genEOM})$,
along with the conditions under which they are relevant, can most
easily be seen after noting that $R = 4 \Lambda_{DE}G/c^2 +8\pi
TG/c^4$, where $T = T_\mu^\mu$. Then $\mathfrak{R}[Rc^2/\Lambda_{DE}
  G]=\mathfrak{R}[4+8\pi T/\Lambda_{DE} c^2]$, where the `4' comes
from the dimensionality of spacetime. It is readily clear that in
regions of spacetime where either $T_{\mu\nu} =0$ or when $T_{\mu\nu}$
is a constant, the right hand side of Eq.~$(\ref{genEOM})$ vanishes,
and our extended GEOM reduces back to the GEOM. 

Beginning with Einstein \cite{Einstein}, there have been a number of
attempts to show that the GEOM are a necessary \textit{consequence} of
the Einstein's equations Eq.~$(\ref{EinsteinEquation})$. Modern
attempts at demonstrating such a linkage \cite{Gero1972, 
  Gero2003} focuses on the energy-momentum tensor, and make the
assumption that the strong energy condition holds: 
$G_{\mu\nu}t^\mu {t'}^\nu \le 0$ (for our signature for the metric),
where $G_{\mu\nu}$ is the Einstein tensor, and $t^\mu, {t'}^\nu$ are
two arbitrary, time-like vectors. They also assume that the background
metric remains fixed during the passage of the test particle. With
this assumption, the background metric decouples from the motion of
the test particle, and can be treated separately. From the dependence
of the extended GEOM on the energy-momentum tensor $T$\textemdash which
includes a contribution from the motion of the test particle
itself\textemdash it is clear that this assumption does not encompass
our extension of the GEOM. It is thus not precluded by \cite{Gero1972,
  Gero2003}. Indeed, we will explicitly construct the energy-momentum
tensor for dust within the framework of the extended GEOM in Section
\textbf{II.D}. 

\subsection{Dynamics of Massless Particles}

For a massless particle, the equations for motion from
$\mathcal{L}_{\hbox{\scriptsize{Ext}}}$ is 
\begin{equation}
v^\nu\nabla_\nu \left(\mathfrak{R}[4+8\pi T/\Lambda_{DE}c^2]v^\mu\right)=0
\label{masslessEOM}
\end{equation}
By reparametizing $dt \to \mathfrak{R}[4+8\pi
  T(x)/\Lambda_{DE}c^2]dt$ \cite{Wald}, Eq.~$(\ref{masslessEOM})$ 
reduces to $v^\nu\nabla_\nu v^\mu=0$. With the correct choice of
parametization, zero-mass particles still obey the GEOM. The
usual general relativistic effects associated with photons\textemdash the 
gravitational redshift and the deflection of light\textemdash
\textit{are thus not effected by our extension of the GEOM.} 
This result is to be expected. Photons are conformal particles, and as
such, do not have an inherent length scale to which effects can be
compared \footnote{Since not all zero-mass particles are
  conformal, this cannot be expected of all such particles. Our
  approach cannot differentiate between conformal and non-conformal
  zero-mass particles.}.   

\subsection{Impact on the Equivalence Principles}

The statements \cite{MTW} of the equivalence principal we are
concerned with here are the following: 

\vskip 10pt
\noindent{\textit{Uniqueness of Free Fall:} It is clear from
  Eq.~$(\ref{genEOM})$ that the worldline of a freely falling test
  particle under the extended GEOM does not depend on its
  composition or structure.

\vskip 10pt
\noindent{\textit{The Weak Equivalence Principle:}  Our extension also
  satisfies the weak equivalence principle to the same level of
  approximation as the GEOM. The weak equivalence principle is
  based on the ability to choose a frame near the worldline
  of the test particle where $\Gamma^\mu_{{\alpha_\Lambda}\beta}
  \approx0$; the Minkowski metric, $\eta_{\mu\nu}$, is thus a good
  approximation to $g_{\mu\nu}$ in the neighborhood around it. However, as one
  deviates from this world line corrections to $\eta_{\mu\nu}$ appear,
  and since a specific coordinate system has been chosen, they appear
  as powers of the Riemann tensor (or its contractions) and its
  derivatives (see \cite{MTW} and \cite{Fermi}). This means that the
  \textit{larger} the curvature, the \textit{smaller} the neighborhood
  about the world line where $\eta_{\mu\nu}$ is a good approximation
  of the metric. Consequently, the weak equivalence principle holds up
  to terms first order in the curvature, and since the additional
  terms in Eq.~$(\ref{genEOM})$ are linear in $R$, our extension of the
  GEOM satisfies the weak equivalence principle to the same
  order of approximation as the GEOM does. 
\vskip 10pt

\noindent{\textit{The Strong Equivalence Principle:} Because we only
  change the geodesic Lagrangian, all nongravitational 
forces in our theory will have the same form as their special relativistic
counterparts. 

\subsection{The Energy-Momentum Tensor}

As we have changed the equations of motion of test particles, we would
expect the energy-momentum tensor for test particles to change as
well. To see how it changes, we begin with the usual tensor for an
inviscid fluid with density $\rho$, pressure $p$, and fluid velocity
$v_\mu(x)$:   
\begin{equation}
T_{\mu\nu} = \rho v_\mu v_\nu -\left(g_{\mu\nu}-\frac{v_\mu
  v_\nu}{c^2}\right) p. 
\label{EM}
\end{equation}
This form for $T_{\mu\nu}$ depends only on the spatial isotropy of the
fluid, and holds for both the GEOM and the extended
GEOM. Following \cite{MTW}, energy and momentum conservation,
$\nabla^\nu T_{\mu\nu}=0$, requires that  
\begin{equation}
0 = v_\nu \nabla^\nu(\rho+p/c^2)v_\mu + (\rho+p/c^2)\nabla_\nu v^\nu v_\mu +
(\rho+p/c^2) v^\nu\nabla_\nu v_\mu - \nabla_\mu p.
\label{Euler}
\end{equation}
Since $v^2 = $ constant even within the extended GEOM
formulation, projecting the above along $v_\mu$ gives once again the
first law of thermodynamics
\begin{equation}
d(V\rho c^2) = - pdV,
\label{thermo}
\end{equation}
where $V$ is the volume of the collection of particles.
\textit{As such, the standard analysis of the evolution of the universe 
follows much in the same way as before under the extended
GEOM.}  

Next, projecting Eq.~$(\ref{Euler})$ long the subspace
perpendicular to $v_\mu$, gives the relativistic version
of Euler's equation 
\begin{equation}
0 = \left(\rho+\frac{p}{c^2}\right) v^\nu \nabla_\nu v_\mu - \left(g_{\mu\nu}-\frac{v_\mu
v_\nu}{c^2}\right)\nabla^\nu p.
\label{spatial}
\end{equation}
We are concerned with the motion of matter in galaxies, and for such a 
system, test particles do not interact with one another except under
gravity. This corresponds to the case of dust. If test particles
in the dust follow the GEOM, then from  
Eq.~$(\ref{spatial})$, $T_{\mu\nu}^{\hbox{\scriptsize{geo-Dust}}} =
\rho v_\mu v_\nu$ and $p\equiv 0$. On the other hand, if the test
particle follow the extended GEOM, the situation changes. Using
Eq.~$(\ref{genEOM})$, Eq.~$(\ref{spatial})$ becomes 
\begin{equation}
0=\left(g_{\mu\nu}-\frac{v_\mu v_\nu}{c^2}\right)\left\{(\rho
c^2+p)\nabla^\nu \log\mathfrak{R} - \nabla^\nu p\right\}, 
\end{equation}
so that 
\begin{equation}
(\rho c^2+p)\nabla_\mu \mathfrak{R} - \mathfrak{R}\nabla_\mu p=\xi
  \Lambda_{DE} cv_\mu,
\label{p}
\end{equation}
where $\xi$ is a constant. By contracting the above with $v_\mu$, it
is straightforward to see that if $\xi\ne0$, $p$ will increase
linearly with the proper time. This would be unphysical, and we
conclude that $\xi$ must be zero. 

As we are interested in the nonrelativistic, linearized
gravity limit, $T=\rho c^2 - 3p \approx \rho 
c^2$; $\mathfrak{R}$ is a function of $\rho$ only in this limit, and so,
consequently, is $p$. Equation~$(\ref{p})$ then results in  
\begin{equation}
p(\rho) = -\rho c^2 + c^2\mathfrak{R}[4+8\pi\rho/\Lambda_{DE}]\int_0^\rho
\frac{ds}{\mathfrak{R}[4+8\pi s/\Lambda_{DE}]}.
\label{pressure}
\end{equation}
Given the density, the pressure\textemdash and thus the
energy-momentum tensor for dust, $T^{\hbox{\scriptsize{Ext-Dust}}}$,
under the extended GEOM\textemdash is determined.

To determine $\rho$, we note from Eq.~$(\ref{pressure})$ that $p\sim
\rho^2$ for $\rho \to 0$, while $p\sim \Lambda_{DE}$ when $\rho\gg
\Lambda_{DE}/8\pi$. We may thus still approximate
$T^{\hbox{\scriptsize{Ext-dust}}}_{\mu\nu} \approx \rho v_\mu v_\nu$
in the nonrelativistic, linearized gravity limit. We next perturb off
the Newtonian metric $\eta_{\mu\nu}$ through $g_{\mu\nu} = 
\eta_{\mu\nu} + h_{\mu\nu}$, where the only nonzero component of
$h_{\mu\nu}$ is $h_{00}=2\Phi/c^2$, and $\Phi$ is the Newtonian
potential. It satisfies 
\begin{equation}
\mathbf{\nabla}^2\Phi + 2\frac{\Lambda_{DE}G}{c^2}\Phi = 4\pi\rho G -
\Lambda_{DE}G, 
\label{PhiEOM}
\end{equation}
in the presence of a cosmological constant. 

As usual, the temporal coordinate, $x^0$, for the extended GEOM is this
limit is approximated by $ct$ to lowest order in $\vert
\mathbf{v}\vert/c$. The spatial coordinates, $\mathbf{x}$, on the
other hand, reduces to   
\begin{equation}
\frac{d^2 \mathbf{x}}{dt^2} = -\mathbf{\nabla} \Phi - \left(\frac{4\pi
     c^2}{\Lambda_{DE}}\right)
     \left[\frac{\mathfrak{D}'(4+8\pi\rho/\Lambda_{DE})}{1+
     \mathfrak{D}\left(4 + 8\pi\rho/\Lambda_{DE}\right)}\right]
     \mathbf{\nabla}\rho.
\label{NREOM}
\end{equation}
In principle, $\rho$ can then be determined through the collection
motion of the stars within galaxies. 

\subsection{A Form for $\mathfrak{D}(x)$ and Experimental Bounds on
  $\alpha_\Lambda$}   

Since our extension of the GEOM does not change the equations of
motion for massless test particles, we expect Eq.~$(\ref{genEOM})$ to
reduce to the GEOM in the ultrarelativistic limit. It is only in the
\textit{nonrelativistic} limit where deviations from geodesic due to
the additional terms in Eq.~$(\ref{genEOM})$ can be seen. We therefore
focus on the impact of the extension in the nonrelativistic,
linearized-gravity limit of Eq.~$(\ref{genEOM})$, and begin by
constructing $\mathfrak{D}(x)$.   

For the addition terms from the extended GEOM \textit{not} to
contribute significantly to Newtonian gravity under current 
experimental conditions, $\mathfrak{D}'(4 + 
8\pi\rho/\Lambda_{DE}) \to 0$ when $\rho >> \Lambda_{DE}/2\pi$. Note
also that in the absence of the additional terms the motion of stars
in galaxies is governed by a Newtonian, $1/r$ potential; what is
instead observed is a weaker, logarithmic potential. These additional
terms in the extended GEOM should thus contribute to the equations of
motion of a test particle as though they were from a repulsive
potential; this requires $\mathfrak{D}'(x)<0$.   

The simplest form for $\mathfrak{D}'(x)$ with these requirements is 
\begin{equation}
\mathfrak{D}'(x) = -\frac{\chi}{1 + x^{1+{\alpha_\Lambda}}},
\end{equation}
where $\chi$ is a normalization constant
\begin{equation}
\frac{1}{\chi} = \int_0^\infty \frac{ds}{1+s^{1+{\alpha_\Lambda}}}.
\end{equation}
To prevent negative effective masses, $\mathfrak{D}(x)$ must be
positive, so that  
\begin{equation}
\mathfrak{D}(x) = \chi(\alpha_\Lambda)\int_x^\infty \frac{ds}{1+s^{1+{\alpha_\Lambda}}},
\end{equation}
where ${\alpha_\Lambda} >0$  for the integral to be defined. While
the precise form of $\mathfrak{D}(x)$ is calculable, we will not need
it. Instead, because $8\pi\rho/\Lambda_{DE}\ge 0$,  
\begin{equation}
\mathfrak{D}(4+8\pi\rho/\Lambda_{DE}) = \chi\sum_{n=0}^\infty
  \frac{(-1)^n}{n(1+{\alpha_\Lambda})+{\alpha_\Lambda}}
  \left(4+\frac{8\pi\rho}{\Lambda_{DE}}\right)^{-n(1+{\alpha_\Lambda})-{\alpha_\Lambda}},
\label{D-expand}
\end{equation}
while
\begin{equation}
\frac{1}{\chi}=1+2\sum_{n=0}^\infty
\frac{(-1)^n}{[1+(n+1)(1+{\alpha_\Lambda})][n(1+{\alpha_\Lambda})+{\alpha_\Lambda}]}.
\end{equation}
Notice that in the ${\alpha_\Lambda}\to\infty$ limit, $\mathfrak{D}(x)\to
0$, $\mathcal{L}_{\hbox{\scriptsize{Ext}}}\to\mathcal{L}_0$, and the
GEOM is recovered. 

Bounds on $\alpha_\Lambda$ will be found below. For now, we note that
for ${\alpha_\Lambda} > 1$, $\chi\sim 1$ and 
$\mathfrak{D}(4+8\pi\rho/\Lambda_{DE}) \approx 0$.  Thus, 
\begin{equation}
\frac{d^2 \mathbf{x}}{dt^2} = -\mathbf{\nabla} \Phi + \left(\frac{4\pi
     c^2\chi}{\Lambda_{DE}}\right)
     \left\{1+\left(4+\frac{8\pi\rho}{\Lambda_{DE}}\right)^{1+{\alpha_\Lambda}}\right\}^{-1}  
     \mathbf{\nabla}\rho. 
\label{FinalEOM}
\end{equation}
From WMAP, $\Lambda_{DE}= 7.21_{-0.84}^{0.82} \times 10^{-30}$
g/cm${}^3$, which for hydrogen atoms corresponds to a number density
of $\sim 4$ atoms/m${}^3$. As such, the density of both solids and
liquids far exceed $\Lambda_{DE}$, and in such media
Eq.~$(\ref{FinalEOM})$ reduces to what one expects for Newtonian
gravity. Only very rare gases, in correspondingly hard vacuums,
can have a density that is so small that the additional terms
are relevant. To see when this may occur, consider the 
hardest vacuum that we know of at $\sim 10^{-13}$ torr
\cite{Ishi}. For a gas of He${}_4$ atoms at 3 deg K,
this corresponds to a density of $\rho_{\hbox{\scriptsize limit}}
\approx 10^{-18}$ g/cm${}^3$. Even  
though $\rho_{\hbox{\scriptsize limit}}$ is still 11 orders of magnitude
smaller than $\Lambda_{DE}$, because the scale of the acceleration
from the additional terms in Eq.~$(\ref{FinalEOM})$ is so
large, effects at these densities can nevertheless be relevant. 

Let us consider an experiment that looks for signatures of the
extension of the GEOM Eq.~$(\ref{FinalEOM})$ by looking for anomalous
accelerations (through pressure fluctuations) in a gas of He${}^4$
atoms at 3 deg K, and $\rho = \rho_{\hbox{\scriptsize limit}}$. Inside
this gas we consider a sound wave with amplitude $\epsilon
\rho_{\hbox{\scriptsize limit}}$ propagating with a wavenumber
$k$. Suppose that the smallest measurable acceleration for a test
particle in this gas is $a_{\hbox{\scriptsize bound}}$. Then, for the
additional terms in Eq.~$(\ref{FinalEOM})$ to be undetectable,
\begin{equation}
a_{\hbox{\scriptsize{bound}}} \ge \frac{c^2\chi}{2}
\left(\frac{\Lambda_{DE}}{8\pi\rho_{\hbox{\scriptsize{limit}}}}\right)^{\alpha_\Lambda} \epsilon k.
\end{equation}
This gives a lower bound on ${\alpha_\Lambda}$ as
\begin{equation}
{\alpha_\Lambda}_{\hbox{\scriptsize{bound}}} = \frac{\log{\left[
      2a_{\hbox{\scriptsize{bound}}}/c^2 \chi\epsilon k
      \right]}}{
    \log{\left[\Lambda_{DE}/8\pi\rho_{\hbox{\scriptsize{limit}}}\right]}}.
\label{alphaBound}
\end{equation}
For $\epsilon =0.1$, $k=1$ cm${}^{-1}$, and
$a_{\hbox{\scriptsize{bound}}} = 1$ cm/s${}^2$, 
${\alpha_\Lambda}_{\hbox{\scriptsize{bound}}}$ ranges from $1.28$ for
$\Lambda_{DE} = 10^{-32}$ g/cm${}^3$ to $1.58$ for $\Lambda_{DE} =
10^{-29}$ g/cm${}^3$. 

In idealized situations such as Einstein's analysis of the
advancement of perihelion of Mercury, the energy-momentum is taken to
be zero outside of a massive body such as the Sun; the right hand term in
Eq.~$(\ref{genEOM})$ will not clearly not affect these analyses. This
argument would seem to hold for all other experimental tests 
of general relativity as well. It is an argument that is too simplistic,
however. In practice, the $T_{\mu\nu}$ in each of these tests does
not, in fact, vanish; there is always a background density 
present. Except for experiments involving electromagnetic 
waves, what is needed instead is a comparison of the background
density with $\Lambda_{DE}$. It is only when this density is much
greater than $\Lambda_{DE}/2\pi$ that the additional terms in
Eq.~$(\ref{genEOM})$ will be negligible.     

We have seen when this condition for the density holds for
terrestrial experiments. Considering now the traditional tests of 
general relativity, only
in experiments involving motion of massive particles\textemdash such as
the motion of Mercury or the state-of-the art E\"ovtos-type
experiments done recently by Adelberger \cite{Adel2001, Adel2003,
  Adel2004}\textemdash will the effects of the extension be seen. The
number density of matter at 
Mercury's orbit is roughly 100 atoms/cm${}^3$, however, corresponding
to a mass density greater than $\sim 10^{-23}$ g/cm${}^3$, which is
orders of magnitude greater than $\Lambda_{DE}$.  It also has a
corresponding length scale, $c/\sqrt{\rho G}$, that is on the order of 12
Mpc, which is orders of magnitude larger than the size of the Solar
System. The additional terms in the extended GEOM thus cannot 
appreciably affect the motion of Mercury, or any other solar body.
Next, while the pressures under which Adelberger's  experiments were
performed were not explicitly stated, as far as we know these
experiments were not done at pressures lower than $10^{-13}$ torr; we
would not expect effects from the additional terms to be apparent in
these experiments either. We therefore would not expect the effects of
Eq.~$(\ref{FinalEOM})$ to have already been seen experimentally. Instead, 
with the average galactic-core density $\sim 10^{-24}
\hbox{\textendash} 10^{-22}$ g/cm${}^{-3}$ and sizes of galaxies $\sim
100$ kpc, it is on the galactic length scales and longer where our
extension to the GEOM will become important, and its effects felt.  

\subsection{Connections with Other Theories}

As unusual as the extended GEOM Eq.~$(\ref{genEOM})$ may appear
to be, there are connections between it and other theories. 

\subsubsection{The Class of Scalar Field Theories in Curved Spacetimes}

The Klein-Gordon equation corresponding to the extended GEOM is 
\begin{equation}
\nabla^2 \phi + \frac{m^2c^2}{\hbar^2}\left(
  1+\mathfrak{D}(4)-4\mathfrak{D}'(4)  +
  \mathfrak{D}'(4)\frac{Rc^2}{\Lambda_{DE} G}
\right)\phi = 0.
\label{extendKG}
\end{equation}
where we have expanded $\mathfrak{D}(Rc^2/\Lambda_{DE}G)$ about $R= 
4\Lambda_{DE}G/c^2$. Although the relativistic Klein-Gordon equation
for a scalar field theory can be straightforwardly generalized to
curved spacetimes, it has also been generalized as
\begin{equation}
\nabla^2\phi_{R} + \left(\frac{m^2c^2}{\hbar^2}+\xi R\right)\phi_R=0,
\label{conformalS}
\end{equation}
since for $\xi = 1/6$ the scalar field will be conformally
invariant even though $m\ne 0$ \cite{BD}. The
similarity between Eqs.~$(\ref{extendKG})$ and $(\ref{conformalS})$ is
readily apparent.

\subsubsection{MOND}

As with MOND, the addition terms in our extension of the
GEOM are nonzero at galactic length scales, while on 
terrestrial or interplanetary scales they are
negligibly small. Like MOND, our extension is able
to explain the galactic rotation curves, as we show in
the next section. Within the MOND theory there is a
fundamental acceleration scale $a_{\hbox{\scriptsize{MOND}}} = cH$
\cite{Sand}; in our analysis the scale that measures the
additional contributions to the GEOM in Eq.~$(\ref{FinalEOM})$
is $a_{\hbox{\scriptsize{Ext}}} \sim c^2/L$. As galactic rotation
curves are driven purely by gravitational effects, $\lambda_{DE}$ is
the only natural length scale, and thus $a_{\hbox{\scriptsize{Ext}}} =
c\sqrt{G\Lambda_{DE}}$; numerically $a_{\hbox{\scriptsize{Ext}}}\sim
a_{\hbox{\scriptsize{MOND}}}$. Our extension of the GEOM
thus gives an explanation for both the modification of Newtonian gravity
that MOND proposes, and the fundamental acceleration scale that
appears in theory. 

However, unlike MOND our extension of the GEOM is done within
the framework of general relativity, and still requires the existence
of Dark Matter. Although at the 
nonrelativistic,  Newtonian level there is no separation between the
force of gravity and the response of matter to it, in general
relativity there is. Our theory, in keeping the form of
Eq.~$(\ref{EinsteinEquation})$, does \textit{not} change how matter
curves spacetime; it changes how spacetime affects matter. Massless
test particles still travel along null-geodesics, and obey the
GEOM; the gravitation redshift, the deflection of light 
by massive objects, and gravitational lensing are all not affected by
our extension. This is not the case for MOND, which was proposed at
the Newtonian gravity level as a new theory of gravity. The theory must not
only be extended to a relativistic one, but the response of
electromagnetic fields must be extended as well, and this extension
must be done in such a way that effects such as the red shift and
gravitational lensing are unchanged. This program is not needed in our
approach.   

\subsubsection{The $f(R)$ Theory}

Proposals for introducing additional terms of the form $f(R)$ to
gravity has been made before (see \cite{Nav-2006-1} and
\cite{Noji-2006} for reviews), but at the level of the Hilbert action
for $R$. These theories were first introduced to explain cosmic
acceleration without the need for Dark Energy \cite{Cap, Turner} using
a $1/R$ action, and further extension have been made \cite{Noji-2003-1,
  Noji-2007}. They are now being studied in their own right, and
various functional forms for $f(R)$ are now being considered. Indeed,
connection to MOND has been made for logarithmic $f(R)$ terms
\cite{Noji-2003-2, Nav-MOND}, and with other choices of $f(R)$,
connection with quintessence has been made \cite{Nav-2006-2, Whit,
  Barr, Maed, Wand} as well. Importantly, issues with the introduction  
of a ``fifth force'', and compatibility with terrestrial experiments
have begun to be addressed through the Chameleon Effect (see
\cite{Khou-1, Khou-2, Brax, Mota} and an overview in
\cite{Nav-2006-2}). This effect is a mechanism for hiding the effects of
field with a small mass that would otherwise be seen.
 
\section{Dark Energy and Galactic Structure}

While definitive, a first principles calculation of the galactic
rotation curves using Eq.~$(\ref{NREOM})$ to describe the motion of
each star in a galaxy would be analytically intractable. Instead, the
approach we will take is to show that \textit{given} a model of a stationary
galaxy with a specific rotation curve, we are able to
\textit{derive} the mass density profile of the galaxy. The logarithmic 
interaction potential observed for the motion of stars in the galactic
disk follows. We then will show that an idealized
pseudoisothermal density profile will result in a lower free-energy
state for the calculated density than an idealized, cuspy-profile
density. 

\subsection{A Model Galaxy}

A number of geometries have been used to model the formation of galaxies
(see \cite{Fill}). Because we will be making connection with
cosmology, we are interested in the large-distance properties of the
density profile, however, and at such distances the detail structures of
galaxies are washed out; only the spherically symmetric features of the
galaxy survive. We thus use a spherical geometry to model our
idealized galaxy, and divide space into the following three
regions. Region I $=\{r \>\>\vert \>\>  r\le r_H \hbox{, and  }\rho
\gg \Lambda_{DE}/2\pi\}$, where $r_H$ is the galactic core radius. Region II
$=\{r \>\>\vert \>\> r> r_H, r \le r_{II}, \hbox{ and  } \rho \gg
\Lambda_{DE}/2\pi\}$ is the region outside the core containing stars
undergoing rotations with constant velocity; it extends out to a
distance of $r_{II}$, which is determined by the theory. A Region III $=\{r
\>\>\vert \>\> r > r_{II} \hbox{, and } \rho \ll \Lambda_{DE}/2\pi\}$
naturally appears in the theory as well. 

We assume that all the stars in the model galaxy undergo circular
motion. While this is an approximation, galactic rotation curves are
determined with stars that undergo such motion, and we use these
curves as inputs for our analysis. The acceleration of each star,
$\mathbf{a} \equiv \ddot{\mathbf{x}}$, is then a function of the 
location, $\mathbf{x}$, of the star only. As such, we can take the
divergence of Eq.~$(\ref{NREOM})$, and obtain  
\begin{equation}
f(\mathbf{r})= \rho - 
\frac{1}{\kappa^2(\rho)} \left\{\mathbf{\nabla}^2\rho -
  \frac{1+{\alpha_\Lambda}}{4+8\pi\rho/\Lambda_{DE}}
  \left(\frac{8\pi}{\Lambda_{DE}}\right) 
  \vert\mathbf{\nabla}\rho\vert^2\right\},
\label{rhoEOM}
\end{equation}
where 
\begin{equation}
\kappa^2(\rho) \equiv
  \frac{1}{\chi\lambda_{DE}^2}\left\{1+\left(4+ 
  \frac{8\pi\rho}{\Lambda_{DE}}\right)^{1+{\alpha_\Lambda}}\right\}, 
\label{kappa}
\end{equation} 
and $f(\mathbf{x}) \equiv -\mathbf{\nabla}\cdot\mathbf{a}/4\pi G$
is considered to be a driving term. 
Because we are dealing with only gravitational forces, we do not
differentiate between baryonic matter and Dark Matter in $\rho$. 

In deriving Eq.~$(\ref{rhoEOM})$, we used the Newtonian relation
$\mathbf{\nabla}^2\Phi = 4\pi\rho G$ instead of the full expression
Eq.~$(\ref{PhiEOM})$. We do so because $\Lambda_{DE}$ is so small that
it may be neglected for most of the regions we are interested
in. While the contribution $2\Lambda_{DE}G\Phi/c^2$ to Eq.~$(\ref{PhiEOM})$
means that $\Phi$ oscillates, it does so on a length scale
$\lambda_{DE}/\sqrt{2}$, which is longer than $\lambda_H/\sqrt{2}$;
we will find that $\rho$ is exponentially small where this scale is
relevant. We are also mostly interested in regions where $\rho >>
\Lambda_{DE}/2\pi$, and in this region the term 
$-\Lambda_{DE}$ on the right-hand-side of Eq.~$(\ref{PhiEOM})$ is
negligible; where $\rho\sim\Lambda_{DE}/2\pi$ is precisely where $\rho \to
0$ exponentially fast. The $\Lambda_{DE}$ term in
Eq.~$(\ref{PhiEOM})$ therefore only insures that $\Phi\to -c^2/2$ as
$r\to\infty$, and this can be taken into account with the appropriate
boundary conditions for $\Phi$.

It is straightforward to see that 
\begin{equation}
\mathbf{\nabla}\cdot \mathbf{a}
=-\frac{1}{r^2}\frac{\partial\>\>}{\partial r}\left[rv(r)^2\right],
\label{div-accel}
\end{equation}
where $v(r)$ is the velocity curve for the galaxy. Thus, given a
$v(r)$, $\mathbf{a}$ can be found and $f(\mathbf{x})$ determined. For
the observed velocity curves, we use a particularly simple
idealization: $v^{\hbox{\scriptsize ideal}}(r) =v_H r/r_H$ for $r \le 
r_H$, while $v^{\hbox{\scriptsize ideal}}(r)=v_H$ for 
$r>r_H$, where $v_H$ is the asymptotic value of the rotation
curve. While $v^{\hbox{\scriptsize ideal}}(r)$ is 
continuous, $f(r)$ is not, and we find 
that $f(r)= 3v_H^2/4\pi G r_H^2 \equiv \rho_H$ for $r\le r_H$ and
$f(r)= v_H^2/4\pi G r^2= \rho_H r_H^2/3r^2$ for $r>r_H$.

Analytically, this idealized velocity curve $v^{\hbox{\scriptsize
    ideal}}(r)$ is more tractable then the velocity curve for the
pseudoisothermal density profile profile \cite{Blok-1}
\begin{equation}
v^{\hbox{\scriptsize p-iso}}(r) =
\sqrt{4\pi G\rho_HR_C^2
  \left[1-\frac{R_C}{r}\arctan{\left(\frac{r}{R_C}\right)}\right]}. 
\label{iso-curve}
\end{equation}
However, because it has the same limiting forms in both the $r\ll r_H$
and $r\gg r_H$ limits, $v^{\hbox{\scriptsize ideal}}(r)$ functions an
idealization of $v^{\hbox{\scriptsize p-iso}}(r)$ as well. For the
$r\gg r_H$ limit, we need only identify $\sqrt{4\pi
  G\rho_HR_C^2}\equiv v_H$, and in the $r\ll r_H $ limit  
we need only identify $r_H = \sqrt{3}R_C$. 

For density profiles with a cusp-like structure Eq.~$(\ref{cusp})$,
the situation is more complicated. Here, it is the density profile
that is given, even though it is the velocity curve that is
observed. While it is possible to integrate the density profile
Eq.~$(\ref{cusp})$ to find the corresponding velocity curves
$v_{\hbox{\scriptsize cusp}}(r)$, both the maximum value 
of $v_{\hbox{\scriptsize cusp}}(r)$ and the point where its slope
changes\textemdash giving the size of the core\textemdash are
different depending on the density profile used. Without a value of
the core size that is consistent from one cusp profile to another, it
is not possible to compare profiles and their free energies. 

Although it is possible in principal to determine the core sizes for
each of the cusp profiles, doing so will be analytically
intractable. Instead, we account for the different density profiles by taking   
\begin{equation}
f(\mathbf{x}) =
\left\{
\begin{array}{ll} 
\rho_H \left(r_H/r\right)^\gamma & \mbox{if $r \le r_H$,}
\\
\frac{1}{3}\rho_H \left(r_H/r\right)^\beta& \mbox{if $r>r_H$.}
\end{array}
\label{drive}
\right.
\end{equation}
The core size is set to be $r_H$, and for the specific case $\gamma=0$ and
$\beta=2$, Eq.~$(\ref{drive})$ reduces to the $f(\mathbf{x})$ for
$v^{\hbox{\scriptsize ideal}}(r)$. For the velocity to be finite at
$r=0$, $\gamma<2$, while for it be finite 
as $r\to\infty$, $\beta \ge2$. We will see that the density profile 
calculated from $f(\mathbf{x})$ goes as $1/r^\gamma$ for $r<r_H$ and
as $ 1/r^\beta$ for $r> r_H$; it is necessarily continuous at
$r=r_H$. Both limiting behaviors are equivalent to
Eq.~$(\ref{cusp})$, and thus Eq.~$(\ref{drive})$ results in an
idealization of the density profiles in Eq.~$(\ref{cusp})$, while
allowing us to consistently compare profiles.  

\subsection{Observational Bounds on ${\alpha_\Lambda}$}

An estimate for ${\alpha_\Lambda}$ can be obtained by comparing two
length scales. From Eq.~$(\ref{rhoEOM})$, the 
density near $r_H$ decreases with characteristic length scale
\begin{equation}
\lambda_{\hbox{\scriptsize{core}}} \approx \frac{1}{\kappa(\rho_H/\Lambda_{DE})} =
  \chi^{1/2}\lambda_{DE}\left(\frac{\Lambda_{DE}}{8\pi
  \rho_H}\right)^{(1+{\alpha_\Lambda})/2}.
\label{lambda_H}
\end{equation}
The size of the galactic core $r_H$ should be proportional to
$\lambda_{\hbox{\scriptsize{core}}} $, since at 
distances much smaller than $r_H$ galactic dynamics 
are driven by Newtonian gravity, while at distances much larger than $r_H$
they are driven by the extended terms in
Eq.~$(\ref{FinalEOM})$. Fixing the $\lambda_{\hbox{\scriptsize{core}}}$ in
Eq.~$(\ref{lambda_H})$, we obtain an estimate for 
\begin{equation}
{\alpha_\Lambda} =
\frac{\log\left(\chi\lambda_{DE}^2/\lambda_{\hbox{\scriptsize{core}}}^2\right)}{\log\left(8\pi\rho_H/\Lambda_{DE}\right)} - 1. 
\label{alpha}
\end{equation}

Figure $\ref{Fig-1}$ shows graphs of ${\alpha_\Lambda}$ in term of
$\Lambda_{DE}$ with the core density fixed at $\rho_H = 10^{-24}$
g/cm${}^3$, and various values of $\lambda_{\hbox{\scriptsize{core}}
}$. The characteristic length $\lambda_{\hbox{\scriptsize{core}}}$
cannot exceed $r_H$, nor can it be too much smaller 
than it; the values of $\lambda_{\hbox{\scriptsize{core}}}$ chosen in
Fig.~1 reflects this. Graphed also is the lower bound on
${\alpha_\Lambda}$ set in section 
\textbf{II.D}. This bound, combined with Eq.~$(\ref{alpha})$,
brackets ${\alpha_\Lambda}$ within the triangle bound by $32
\hbox{\scriptsize{ pc}}\le \lambda_{\hbox{\scriptsize{core}}} \le 100$
pc and $\Lambda_{DE} \ge 3.3 \times 10^{-31}$ g/cm${}^3$, and limits
${\alpha_\Lambda} \ge 1.35$; $\Lambda_{DE} = 7.21 \times 10^{-30}$
g/cm${}^3$ lies within this triangle. Given this result,
we take ${\alpha_\Lambda} = 3/2$ as the representative value of
${\alpha_\Lambda}$ for this part of the paper. A definitive value for
$\alpha_\Lambda$ will be set in Section \textbf{V}. 

\begin{figure}
\begin{center}
\hspace{-1in}
\includegraphics[width=0.7\textwidth, angle = 0]{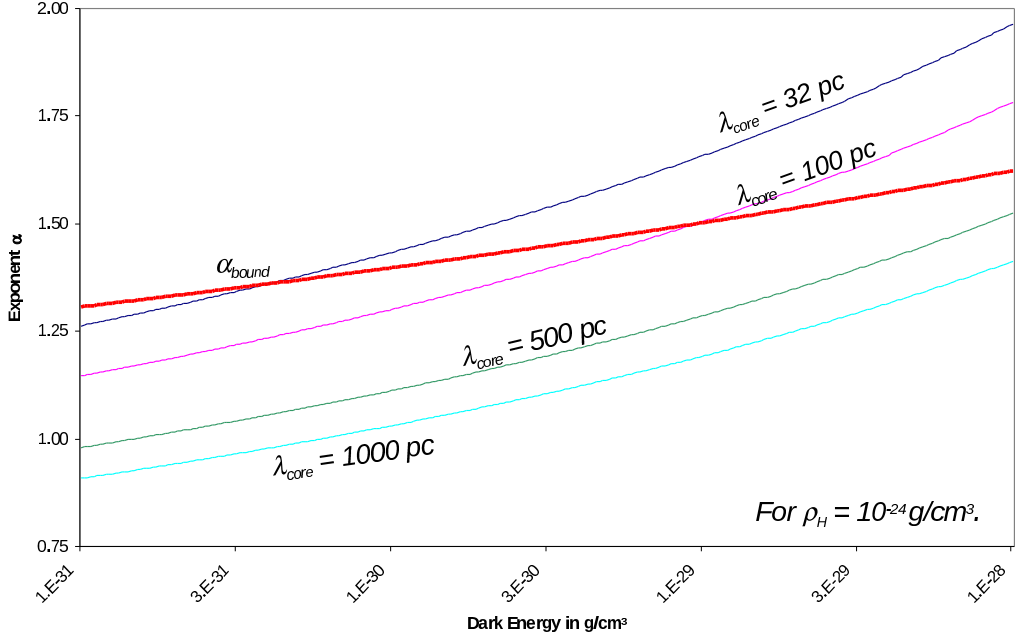}
\end{center}
\caption{\label{Fig-1}
Graphs of ${\alpha_\Lambda}$ with respect to $\Lambda_{DE}$ at $\rho_H =
10^{-24}$ g/cm${}^3$ for various values of
$\lambda_{\hbox{\scriptsize{core}}}\le r_H$. The graph of the lower
experimental bound Eq.~$(\ref{alphaBound})$ is also included. 
}
\end{figure}

\section{The Density Profile of the Model Galaxy}

In both Regions I and II, $\rho_H\gg \Lambda_{DE}/2\pi$ and
Eq.~$(\ref{rhoEOM})$ may be approximated as the following, 
\begin{equation}
f(\mathbf{{u}}) = \rho - \left(\frac{\Lambda_{DE}}{8\pi\rho}\right)^{1+\alpha_\Lambda} \left\{\mathbf{\nabla}_{u}^2\rho - 
  (1+\alpha_\Lambda)\frac{\vert\mathbf{\nabla}_{u}\rho\vert^2}{\rho}\right\},
\label{rhoI-II-EOM}
\end{equation}
where ${u} = r/\chi^{1/2}\lambda_{DE}$, and $\mathbf{\nabla}_{u}$
denotes derivative write respect to ${u}$. The solution to
Eq.~$(\ref{rhoI-II-EOM})$ minimizes the following functional of the
density:  
\begin{eqnarray}
\mathcal{F}[\rho] =
\frac{\Lambda_{DE}c^2}{8\pi}\left(\chi^{1/2}\lambda_{DE}\right)^3 \int 
d^3\mathbf{{u}}
&{}&
\Bigg\{ 
     \frac{1}{2\alpha_\Lambda}
     \Bigg\vert \mathbf{\nabla}_{u} 
          \left(\frac{\Lambda_{DE}}{8\pi\rho}\right)^{\alpha_\Lambda}
     \Bigg\vert^2  
-
\frac{\alpha_\Lambda}{\alpha_\Lambda-1}
\left(\frac{\Lambda_{DE}}{8\pi\rho}\right)^{\alpha_\Lambda-1}
+
\nonumber
\\
&{}&
\left(\frac{\Lambda_{DE}}{8\pi\rho}\right)^{\alpha_\Lambda} \frac{8\pi
  f({u})}{\Lambda_{DE}}\Bigg\},
\label{free-energy}
\end{eqnarray}
which we view as an effective free energy for the system. Here,
we have chosen the scale as $\Lambda_{DE}c^2$. 

In Region III, on the other hand, the following linearization of 
Eq.~$(\ref{rhoEOM})$ is appropriate,
\begin{equation}
0 = \rho -
\frac{1}{1+4^{1+{\alpha_\Lambda}}}\mathbf{\nabla}^2_{u}\rho. 
\label{rhoIII-EOM}
\end{equation}
We will find from a detailed analysis of the solution
for $\rho$ in Region II that the driving term $f(\mathbf{r})$ is
negligibly small in Region III. A calculation of the free energy
in this region will not be necessary. 

\subsection{The Solution for Region I}

For the $v^{\hbox{\scriptsize ideal}}(r)$ curve\textemdash corresponding to
$\gamma=0$\textemdash it is clear that the only solution, $\rho_I(r)$, for
Eq.~$(\ref{rhoI-II-EOM})$ in Region I is the constant solution
$\rho_I(r) = \rho_H$. The free energy for this solution is easily
calculated  
\begin{equation}
{}^I\mathcal{F}_{\gamma=0} =
-\frac{1}{\alpha_\Lambda-1}\frac{\Lambda_{DE}r_H^3}{6}
\left(
     \frac{\Lambda_{DE}}{8\pi\rho_H}
\right)^{\alpha_\Lambda-1}.
\end{equation}

For a general $\gamma > 0$, perturbation theory is used to find
solutions Eq.~$(\ref{rhoI-II-EOM})$. We first scale $\mathfrak{y} =
\rho/\rho_H$ and $\mathfrak{x} = u/u_H \le 1$, so that
\begin{equation}
\mathfrak{x}^\gamma=\mathfrak{y}
+\frac{\varepsilon}{\alpha_\Lambda}\nabla^2\mathfrak{y}^{-\alpha_\Lambda},
\label{epsilon}
\end{equation}
where the small parameter
\begin{equation}
\varepsilon =
\frac{1}{u_H^2}\left(\frac{\Lambda_{DE}}{8\pi\rho_H}\right)^{1+\alpha_\Lambda}\sim
10^{-2} \hbox{\textendash} 10^{-9},
\end{equation} 
for $\alpha = 3/2$, $\rho_H \sim 10^{-24} \hbox{\textendash} 10^{-22}$
g/cm$^3$, and $r_H \sim 1$ kpc \textendash $\>8$ kpc. There are two
approaches to solving Eq.~$(\ref{epsilon})$ perturbatively. The first
treats the $\nabla^2$ term as a perturbation on the solution $\rho_I =
f$. Doing so gives 
\begin{equation}
\rho_I^{(a)}(r) = \rho_H
\left(\frac{r}{r_H}\right)^{-\gamma}\left[1-
\varepsilon\gamma(1+\gamma\alpha_\Lambda)  
\left(\frac{r}{r_H}\right)^{\gamma(1+\alpha_\lambda)-2}\right].
\end{equation}
It is valid when $\gamma \ge 2/(1+\alpha_\lambda)$. For the second, we
take $\mathfrak{z}= \varepsilon/\mathfrak{y}^\alpha_\Lambda$, so that 
\begin{equation}
\mathfrak{x}^{-\gamma}=\varepsilon^{1/\alpha_\Lambda}
\mathfrak{z}^{-1/\alpha_\Lambda}
+\frac{1}{\alpha_\Lambda}\nabla^2\mathfrak{z}.
\end{equation} 
Treating
$\varepsilon^{1/\alpha_\Lambda}\mathfrak{z}^{-1/\alpha_\Lambda}$ now
as the perturbation, the solution for $\mathfrak{z}$ that is 
finite at $\mathfrak{x}=0$ gives
\begin{eqnarray}
\rho_I^{(b)}(r) &=&
\left(\frac{\varepsilon(2-\gamma)(3-\gamma)}{\alpha}\right)^{1/\alpha_\Lambda}
\left(\frac{r}{r_H}\right)^{-\widetilde{\gamma}}
\nonumber
\\
&{}&
\left(1+
\frac{1}{\alpha}
\left[\frac{\varepsilon(2-\gamma)(3-\gamma)}{\alpha}\right]^{1/\alpha_\Lambda}
\frac{(2-\gamma)(3-\gamma)}{(2-\widetilde\gamma)(3-\widetilde\gamma)}
\left(\frac{r}{r_H}\right)^{2-\widetilde\gamma(1+\alpha_{\Lambda})}\right),
\end{eqnarray}
where $\widetilde\gamma = (2-\gamma)/\alpha_\Lambda$. The solution is
now valid for $\widetilde\gamma \le2/(1+\alpha_\Lambda)$, which again
corresponds to $\gamma \ge 2/(1+\alpha_\Lambda)$.

Calculating the free energy for these two perturbative solutions follows
straightforwardly. To lowest order in $\varepsilon$, the free energy
for $\rho_I^{(a)}$ is 
\begin{equation}
{}^I\mathcal{F}_{\gamma> 0}^{(a)} =
-\frac{1}{\alpha_\Lambda-1}\frac{\Lambda_{DE} c^2 r_H^3}{6}
  \left(\frac{\Lambda}{8\pi\rho_H}\right)^{\alpha_\Lambda -1}
  \frac{1}{1+ \gamma(\alpha_\Lambda -1)/3}, 
\end{equation}
while the free energy for $\rho_I^{(b)}$ is
\begin{equation}
{}^I\mathcal{F}_{\gamma> 0}^{b} =
\frac{\alpha\Lambda c^2
  r_H^3}{(2-\gamma)(3-\gamma)^2}\left(\frac{8\pi\rho_H
    u_H}{\Lambda_{DE}}\right)^2. 
\end{equation}

It is clear that ${}^I\mathcal{F}_{\gamma=0} <
{}^I\mathcal{F}^{(a)}_{\gamma\ge 2/(1+\alpha_\Lambda)} < 0 < {}^I\mathcal{F}^{(b)}_{\widetilde{\gamma}<2/(1+\alpha_\Lambda)}$.
The idealized, pseudoisothermal profile corresponding to
$\gamma=0$ is thus the state of lowest free energy in Region I.
Physically, this results because of the curvature term $\sim
\vert\nabla \rho\vert^2 \ge0$ in Eq.~$(\ref{free-energy})$. Just like
the lowest free energy state of a Landau-Ginzberg free energy
functional, this term only vanishes for the constant solution; for all
other solutions it contributes positively to the free
energy. 

\subsection{The Solution for Region II}  

In this region, 
\begin{equation}
\frac{1}{3}\rho_H \left(\frac{r_H}{r}\right)^\beta = \rho_{II} - 
 \chi \lambda_{DE}^2
 \left(\frac{\Lambda_{DE}}{8\pi\rho_{II}}\right)^{1+{\alpha_\Lambda}} \left\{
 \mathbf{\nabla}^2 \rho_{II} -
 (1+{\alpha_\Lambda})\frac{\vert\mathbf{\nabla}\rho_{II}\vert^2}{\rho_{II}}\right\},
\label{RII-EOM}
\end{equation}
and we denote $\rho_{II}$ as the solution for $\rho$ in Region
II. We undertake an asymptotic analysis \cite{Bender} of
Eq.~$(\ref{RII-EOM})$ by making the anzatz that
within Region II there exists a point $r_{\hbox{\scriptsize{asymp}}}$
beyond which $\rho_H (r_H/r)^\beta/3 \ll \rho(r)$. For
$r>r_{\hbox{\scriptsize{asymp}}}$, we can then
neglect the driving term in Eq.~$(\ref{RII-EOM})$, leaving the
homogeneous equation 
\begin{equation}
0 = \rho_{\hbox{\scriptsize{asymp}}} -
 \left(\frac{\Lambda_{DE}}{8\pi\rho_{\hbox{\scriptsize{asymp}}}}\right)^{1+{\alpha_\Lambda}} 
 \left\{\mathbf{\nabla}_{{u}}^2 \rho_{\hbox{\scriptsize{asymp}}} -
 (1+{\alpha_\Lambda})\frac{\vert\mathbf{\nabla}_{{u}}
 \rho_{\hbox{\scriptsize{asymp}}}\vert^2}{\rho_{\hbox{\scriptsize{asymp}}}}\right\}. 
\label{asympEOM}
\end{equation}

\subsubsection{Asymptotic Analysis and the Background Density}

We look for a power-law solution to Eq.~$(\ref{asympEOM})$ with the form
\begin{equation}
\rho_{\hbox{\scriptsize{asymp}}} = \frac{\Lambda_{DE}}{8\pi} \Sigma({\alpha_\Lambda}) {u}^p,
\end{equation}
and find that 
\begin{equation}
0=-1+\frac{p(1-{\alpha_\Lambda} p)}{\bigg(\Sigma({\alpha_\Lambda})\bigg)^{1+{\alpha_\Lambda}}}
\frac{1}{u^{p(1+{\alpha_\Lambda})+2}}.
\label{asympConstraint}
\end{equation}
This gives $p=-2/(1+{\alpha_\Lambda})$, with $\Sigma(\alpha_\Lambda)$ 
the solution of  
\begin{equation}
0=1+\frac{2(1+3{\alpha_\Lambda})}{(1+{\alpha_\Lambda})^2
  \big[\Sigma({\alpha_\Lambda})\big]^{1+{\alpha_\Lambda}}}.
\label{amp}
\end{equation}
Positivity of $\rho_{\hbox{\scriptsize{asymp}}}$ requires that $\Sigma({\alpha_\Lambda})
>0$, and this requires that there are positive solutions to
Eq.~$(\ref{amp})$. Such solutions exist only if ${\alpha_\Lambda}$ is the
ratio of a odd integer with an even integer. Our choice of
${\alpha_\Lambda} = 3/2$ satisfies this criteria, and we arrive at the
following asymptotic solution
\begin{equation}
\rho_{\hbox{\scriptsize{asymp}}} = \frac{\Lambda_{DE}}{8\pi} \Sigma({\alpha_\Lambda})
\left(\frac{\chi\lambda_{DE}^2}{r^2}\right)^{1/(1+{\alpha_\Lambda})},
\end{equation}
where
\begin{equation}
\Sigma({\alpha_\Lambda}) =
\left[\frac{2(1+3{\alpha_\Lambda})}{(1+{\alpha_\Lambda})^2}\right]^{1/(1+{\alpha_\Lambda})}.
\label{Sigma}
\end{equation}

To justify our anzatz that $r_{\hbox{\scriptsize{asymp}}}$ lies within
Region II, we set $f(r_{\hbox{\scriptsize{asymp}}})=
\rho_{\hbox{\scriptsize{asymp}}}(r_{\hbox{\scriptsize{asymp}}})$, and find
\begin{equation}
\frac{r_{\hbox{\scriptsize{asymp}}}}{r_H} \le
\left(\frac{8\pi\rho_H}{3\Lambda_{DE}}
\frac{u_H^{2/(1+\alpha_\Lambda)}}
     {\Sigma(\alpha_\Lambda)}\right)^{(1+\alpha_\Lambda)/2\alpha_\Lambda},
\end{equation}
where $u_H = r_H/\chi^{1/2}\lambda_{DE}$. For ${\alpha_\Lambda} =
3/2$, $\rho_H\sim 10^{-24}$ g/cm${}^3$, and $r_H=1$ kpc,
$r_{\hbox{\scriptsize{asymp}}}\le 1.77 r_H$; the anzatz is   
valid through the great majority of Region II for the range of
galaxies we are interested in. The upper limit $r_{II}$ to Region II,
on the other hand, is found by setting 
$8\pi\rho_{\hbox{\scriptsize{asymp}}}(r_{II})/\Lambda_{DE}=4$, which gives
\begin{equation}
r_{II} =
\left[\frac{1}{4}\Sigma({\alpha_\Lambda})\right]^{(1+{\alpha_\Lambda})/2}
\chi^{1/2}\lambda_{DE}. 
\end{equation}
For ${\alpha_\Lambda} = 3/2$, $r_{II} \approx 0.20\>\lambda_{DE}$.

\subsubsection{The Near Core Density}

Structural details of the galaxy cannot be seen from
$\rho_{\hbox{\scriptsize{asymp}}}$. Instead, we take
$\rho_{II} = \rho_{\hbox{\scriptsize{asymp}}}+\rho_{II}^{1}$ and expand
Eq.~$(\ref{RII-EOM})$ to first order in $\rho_{II}^{1}$, 
\begin{equation}
\frac{2(1+3{\alpha_\Lambda})}{(1+{\alpha_\Lambda})^2}
\left[\frac{\rho_H}{3}\left(\frac{{u}_H}{{u}}\right)^\beta \right]=
\frac{2(1+3{\alpha_\Lambda})}{(1+\alpha_\Lambda)^2}
\frac{\widehat{\rho}_{II}^{\>1}}{{u}^2} +
\mathbf{\nabla}^2_{u} \widehat{\rho}^{\>1}_{II}.
\label{rho1-EOM}
\end{equation}
where $\widehat{\rho}^{\>1}_{II} ={u}^2\rho^{1}_{II} $. In
the special case $\beta =2$, the particular solution to 
Eq.~$(\ref{rho1-EOM})$ is again the constant solution, but now for
$\widehat{\rho}^{\>1}_{II}$; this corresponds to
$\rho^{1}_{II-\beta=2} = \rho_H
r_H^2/3r^2$, as expected for an idealized pseudoisothermal profile.

Boundary conditions for $\widehat{\rho}_{II}^{\>1}$ are set at the
$r=r_H$ surface, 
\begin{equation}
\rho_H = \rho_{\hbox{\scriptsize{asymp}}}(r_H) +
\frac{\widehat{\rho}_{II}^{\>1}(r_H)}{{u}_H^2}, \qquad 
0 =
\frac{\partial\rho_{\hbox{\scriptsize{asymp}}}}{\partial{u}}\Bigg\vert_{u_H}
+
\frac{\partial\>\>\>}{\partial{u}}
\left(\frac{\widehat{\rho}^{\>1}_{II}}{{u}^2}\right)\Bigg\vert_{u_H},
 \label{BC}
\end{equation}
where we have made use of the result of the Region I free energy
analysis and set $\gamma=0$. The solution to Eq.~$(\ref{rho1-EOM})$
for these boundary conditions is
\begin{equation}
\rho_{II}^{1} = \frac{1}{3}
A_\beta\rho_H\left(\frac{r_H}{r}\right)^\beta + 
\left(\frac{r_H}{r}\right)^{5/2}
\left(C_{\cos}\cos\left[\nu_0\log\left(\frac{r}{r_H}\right)\right] +
  C_{sin}\sin\left[\nu_0\log\left(\frac{r}{r_H}\right)\right]\right),
\label{rho-beta}
\end{equation}
where $\nu_0 = \left[2(1+3\alpha_\Lambda)/(1+\alpha_\Lambda)^2 -
  1/4\right]^{1/2}$, and 
\begin{eqnarray}
A_\beta &=& 
\frac{\nu_0^2+1/4}{\nu_0^2 + (5/2 - \beta)^2},
\nonumber
\\
C_{\cos} &=& \rho_H
-\frac{1}{3}A_\beta\rho_H
-\frac{\Lambda_{DE}}{8\pi}\frac{\Sigma(\alpha_\Lambda)}{
{{u}_H}^{2/(1+\alpha_\Lambda)}},
\nonumber
\\
\nu_0 C_{\sin} &=& \frac{5}{2}\rho_H
-\frac{1}{3}A_\beta\rho_H(5/2-\beta)
-\frac{1}{2}\frac{(1+5\alpha_\Lambda)}{(1+\alpha_\Lambda)}
  \frac{\Lambda_{DE}}{8\pi}\frac{\Sigma(\alpha_\Lambda)}{ 
{{u}_H}^{2/(1+\alpha_\Lambda)}}.
\end{eqnarray}

The density $\rho_{II}(r)$ thus consists of the sum of two parts. The
first part, $\rho_{\hbox{\scriptsize{asymp}}}(r)$, corresponds to the
background density, and depends solely on Dark Energy, fundamental
constants, the exponent $\alpha_\Lambda$, and the dimensionality and
underlying spatial symmetry of the spacetime. \textit{It is
universal, and has the same form irrespective of the detailed
  structure of the galaxy}. The second part, $\rho_{II}^{1}(r)$,
\textit{does} depend on the detail structure of the 
galaxy. Variation in $\rho_{II}^{1}$ are
measured on a scale set by $r_H$, the core size, in contrast to
$\rho_{\hbox{\scriptsize{asymp}}}$, whose variations are measured on
a scale set by $\lambda_{DE}$, the Dark Energy length scale.
While our analysis is done only to first order in the
perturbation of $\rho_{II}$, this feature of $\rho_{II}^{1}$ holds
to higher orders as well. 

The perturbation, $\rho_{II}^{1}$, itself depends on two terms. The first
has a power law dependence of $(r_H/r)^\beta$, while the second has a
power law dependence of $(r_H/r)^{5/2}$. Thus, near the galactic
core $\rho_{II}^{1} \sim r^{-q}$, where $q = \hbox{max}(\beta,
5/2)$; to this level of approximation, the density profile near the core
varies at least as fast as $1/r^{5/2}$, irrespective of whether the
density profile is cuspy or pseudoisothermal. For the $r\gg r_H$ the opposite is true, and now $q
= \hbox{min}(\beta,5/2)$; $\rho_{II}^{1}$ decreases no faster than
$1/r^{5/2}$. 
  
\subsubsection{Free Energy Analysis}

The free energy for the density $\rho_{II}$ separates into three terms: 
${}^{II}\mathcal{F} ={}^{II}\mathcal{F}_{\hbox{\scriptsize{asymp}}} +
{}^{II}\mathcal{F}_{\hbox{\scriptsize{asymp}}-\beta} + {}^{II}\mathcal{F}^{1}$,
where
\begin{eqnarray}
{}^{II}\mathcal{F}_{\hbox{\scriptsize{asymp}}} &\equiv& 
\frac{\Lambda_{DE}c^2}{8\pi}\left(\chi^{1/2}\lambda_{DE}\right)^3 \int_{D_{II}}
d^3\mathbf{{u}}\Bigg\{ 
     \frac{1}{2\alpha_\Lambda}
     \Bigg\vert \mathbf{\nabla}_{u} 
          \left(\frac{\Lambda_{DE}}{8\pi\rho_{\hbox{\scriptsize{asymp}}}}\right)^{\alpha_\Lambda} 
     \Bigg\vert^2  
-
\nonumber
\\
&{}&
\frac{\alpha_\Lambda}{\alpha_\Lambda-1}
\left(\frac{\Lambda_{DE}}{8\pi\rho_{\hbox{\scriptsize{asymp}}}}\right)^{\alpha_\Lambda-1}
\Bigg\},
\nonumber
\\
&=& 
\frac{4\alpha_\Lambda^2\Lambda_{DE}c^2 (\chi^{1/2}\lambda_{DE})^{3/2}
\left[\Sigma(\alpha_\Lambda)\right]^2}
     {(\alpha_\Lambda^2-1)(1+5\alpha_\Lambda)\left[\Sigma(\alpha_\Lambda)\right]^{2(1+\alpha_\Lambda)}} 
{u}_{II}^{(1+5\alpha_\Lambda)/(1+\alpha_\Lambda)},
\end{eqnarray} 
is the contribution to the free energy due to the background density
only. The integration is over Region II, and ${u}_{II} =
r_{II}/\chi^{1/2}\lambda_{DE}$.  

The second term is
\begin{eqnarray}
\frac{{}^{II}\mathcal{F}_{\hbox{\scriptsize{asymp}}-\beta}}
     {\left(\chi^{1/2}\lambda_{DE}\right)^3} &\equiv&  
c^2\int_{D_{II}} d^3\mathbf{{u}}f(u) 
\left(\frac{\Lambda_{DE}}{8\pi\rho_{\hbox{\scriptsize{asymp}}}}\right)^{\alpha_\Lambda}+
\nonumber
\\
&{}&
\frac{8\pi\alpha_\Lambda c^2}
     {\Lambda_{DE}[\Sigma(\alpha_\Lambda)]^{2(1+\alpha_\Lambda)}}\int_{\partial  
  D_{II}}
u^4\rho_{II}^{1}(u)\mathbf{\nabla}\rho_{\hbox{\scriptsize{asymp}}}\cdot
d\mathbf{S}.
\nonumber
\\
&=& \frac{\alpha_\Lambda (8\pi)^2 c^2}{\Lambda_{DE}(1+\alpha_\Lambda) 
  [\Sigma(\alpha_\Lambda)]^{2(1+\alpha_\Lambda)} }
\left\{u_H^5\rho_{II}^{1}(u_H) \rho_{\hbox{\scriptsize{asymp}}}(u_H)
-u_{II}^5\rho_{II}^{1}(u_{II}) \rho_{\hbox{\scriptsize{asymp}}}(u_{II})\right\}
\nonumber
\\
&{}&-\frac{\Lambda_{DE}c^2(1+\alpha_\Lambda)^2}{2(1+3\alpha_\Lambda)}
\left(\frac{4\pi\rho_H}{3\Lambda_{DE}}\right) \Sigma(\alpha_\Lambda)u_H^\beta
  \left(\frac{u_{II}^{5-2/(1+\alpha_\Lambda)-\beta}-u_{H}^{5-2/(1+\alpha_\Lambda-\beta)}}
       {5-2/(1+\alpha_\Lambda)-\beta}\right),
\end{eqnarray} 
where $\partial D_{II}$ is the boundary of $D_{II}$ at $r=r_H$ and
$r=r_{II}$. This is the
contribution to the free energy due to the interaction between
$\rho_{\hbox{\scriptsize asymp}}$ and $f(\mathbf{r})$. It is
straightforward to see that
\begin{equation}
{}^{II}\mathcal{F}_{\hbox{\scriptsize{asymp}}-\beta}\sim
\left\{
\begin{array}{ll} 
-(u_H/u_{II})^\beta & \mbox{if $\beta < 5/2$,}
\\
-(u_H/u_{II})^{5/2}& \mbox{if $5/2 \le\beta<5-2/(1+\alpha_\Lambda)$.}
\\
\pm u_H^{5-2/(1+\alpha_\Lambda)}& \mbox{if
  $5-2/(1+\alpha_\Lambda)\le\beta$,} 
\end{array}
\right.
\end{equation}
where the sign of the last term depends on the values of $\beta,
\alpha_\Lambda, u_H$ and $\rho_H$. The magnitude of this term is very
small, however, and it is clear that
${}^{II}\mathcal{F}_{\hbox{\scriptsize{asymp}}-\beta=0}<
{}^{II}\mathcal{F}_{\hbox{\scriptsize{asymp}}-\beta<5-2/(1+\alpha_\Lambda)}< 
{}^{II}\mathcal{F}_{\hbox{\scriptsize{asymp}}-\beta>5-2/(1+\alpha_\Lambda)}$.

The third term
\begin{eqnarray}
{}^{II}\mathcal{F}^{1} &\equiv& 
-\frac{\alpha_\Lambda(1+\alpha_\Lambda)c^2\left(\chi^{1/2}\lambda_{DE}\right)^3}{2[\Sigma(\alpha_\Lambda)]^{2(1+\alpha_\Lambda)}}
\int_{\partial D_{II}}u^4\rho_{II}^{\>1}(u)
\frac{\mathbf{\nabla}\rho_{\hbox{\scriptsize{asymp}}}}
     {\rho_{\hbox{\scriptsize{asymp}}}}\cdot d\mathbf{S}
+ \frac{\alpha_\Lambda\left(\chi^{1/2}\lambda_{DE}\right)^3}{[\Sigma(\alpha_\Lambda)]^{2(1+\alpha_\Lambda)}}
\left(\frac{8\pi c^2}{\Lambda_{DE}}\right) 
\nonumber
\\
&{}&
\int_{D_{II}}
d^3\mathbf{{u}}\Bigg\{
     \frac{1}{2}
     \left\vert \mathbf{\nabla}_{u}\widehat{\rho}_{II}^{\>1} \right\vert^2  
-
\frac{(1+3\alpha_\Lambda)}{(1+\alpha_\Lambda)^2}
\frac{(\widehat{\rho}_{II}^{\>1})^2}{{u}^2}
+
\frac{2(1+3\alpha_\Lambda)}{(1+\alpha_\Lambda)^2}f(u)\widehat{\rho}_{II}^{\>1}
\Bigg\}
\end{eqnarray} 
is the contribution to the free energy due to $\rho_{II}^{1}$ only.
As it $\sim (\rho_{II}^{1})^2$, this term is very
small compared to the other two terms that make up the free 
energy, and can be neglected. We only note that like
${}^{I}\mathcal{F}$, it is the constant solution to the differential
equation that gives the lowest value of ${}^{II}\mathcal{F}^{1}$,
but now the equation is for $\widehat{\rho}_{II}^{\>1} =
u^2\rho_{II}^{1}$, not $\rho_{II}^{1}$. This once again
corresponds to $\beta = 2$, and as such, to a $\rho_{II}^{1}\sim 1/r^2$.

The total free energy, ${}^{II}\mathcal{F}$, in this region is thus
smaller for $\beta = 2 $ than for $\beta >2$. 
Combined with the calculation for ${}^{I}\mathcal{F}$, we conclude
that the pseuodoisothermal rotational velocity curve will result in a density
profile that gives the lowest free energy, and is the preferred state
of the system. Other rotational velocity curves will result in
density profiles that have a higher free energy. We therefore take
$\gamma=0$ and $\beta=2$ for the rest of this paper. 

\subsection{The Solution for Region III}

The solution to Eq.~$(\ref{rhoIII-EOM})$ follows using standard methods,
with the boundary condition $\rho_{II}(r_{II}) = \rho_{III}(r_{II})$.
The length scale
$\sqrt{\chi/(1+4^{1+{\alpha_\Lambda}}})\lambda_{DE}\approx 0.15
\>\lambda_{DE}$ for ${\alpha_\Lambda} = 3/2$, while $r_{II} \approx
0.20\>\lambda_{DE}$; to a good approximation $r_{II} \approx
\sqrt{\chi/(1+4^{1+{\alpha_\Lambda}})}\lambda_{DE}$. As
$\sqrt{\chi/(1+4^{1+{\alpha_\Lambda}}})\lambda_{DE}$ is a scale set by
the theory, we shall use this last expression for $r_{II}$ from now
on. Note also that Regions II and III overlap, and our approach
of solving the nonlinear partial differential equation is self-consistent. 

The only solution that is spherically symmetric and finite at
$r\to\infty$ is
\begin{equation}
\rho_{III}(r) = \frac{\Lambda_{DE}}{8\pi}\Sigma({\alpha_\Lambda})
\frac{\sqrt{\chi}\lambda_{DE}}{r}
\left(1+4^{1+{\alpha_\Lambda}}\right)^{\frac{1}{2}(1-{\alpha_\Lambda})/(1+{\alpha_\Lambda})}
\exp\left(1-\frac{r}{\lambda_{DE}}
\sqrt{
     \frac{1+4^{1+{\alpha_\Lambda}}}{\chi}
}\right),
\label{rho3-solution}
\end{equation}
and in this region the density decreases exponentially fast. 

\subsection{The Effective Potential}

Note that Eq.~$(\ref{NREOM})$ may be written in terms of an effective
potential, $\mathfrak{V}(\mathbf{x})$, as $\ddot{\mathbf{x}} = -
\mathbf{\nabla}\mathfrak{V}$ where
\begin{equation} 
\mathfrak{V}(\mathbf{x}) = \Phi(\mathbf{x}) +
c^2\log\mathfrak{R}[4+8\pi\rho/\Lambda_{DE}].
\label{effPot}
\end{equation}
It is thus \textit{not} the gravitational potential
$\Phi(\mathbf{x})$ that determines the dynamics, but
$\mathfrak{V}(\mathbf{x})$. To see the implications of this, 
we begin by calculating $\Phi(\mathbf{x})$, but only for
Regions I and II. In Region III, $r>r_{II}$, and motion in this
region is unphysical. 

Integrating $\mathbf{\nabla}^2\Phi = 4\pi G\rho$ gives in Region I
\begin{equation}
\Phi(r) = \frac{1}{2}v_H^2 \left(\frac{r}{r_H}\right)^2 -\frac{c^2}{2},
\end{equation}
where the $c^2$ term comes from requiring $\Phi(r_{II})
= -c^2/2$ instead of $\Phi(\infty)=0$. That this is the usual
expression for the Newtonian potential can be 
seen from $\rho_H=3v_H^2/4\pi Gr_H^2$. For Region II, 
\begin{eqnarray}
\Phi(r) = &{}&
\frac{c^2}{4\alpha_\Lambda}
\frac{\chi\Sigma(\alpha_\Lambda)(1+\alpha_\Lambda)^2}{(1+3\alpha_\Lambda)}
{u}^{2\alpha_\Lambda/(1+\alpha_\Lambda)} -2\pi Gr^2\rho_{II}^{1}(r)
\frac{(1+\alpha_\Lambda)^2}{(1+3\alpha_\Lambda)}+
v_H^2\log\left(\frac{r}{r_h}\right)+ 
\nonumber
\\
&{}& -\frac{c^2}{2}+
v_H^2\left[1+4\frac{(1+\alpha_\Lambda)^2}{(1+3\alpha_\Lambda)}\right] 
-\frac{1}{4}
c^2\chi\Sigma(\alpha_\Lambda)
\frac{(3+5\alpha_\Lambda)(1+\alpha_\Lambda)}{(1+3\alpha_\Lambda)}
u_H^{2\alpha_\Lambda/(1+\alpha_\Lambda)}  -
\nonumber
\\
&{}&
\left\{6v_H^2\frac{(1+\alpha_\Lambda)^2}{(1+3\alpha_\Lambda)}
-\frac{c^2}{2}\chi\Sigma(\alpha_\Lambda)\frac{(1+\alpha_\Lambda)(1+2\alpha_\Lambda)}{(1+3\alpha_\Lambda)}u_H^{2\alpha_\Lambda/(1+\alpha_\Lambda)}\right\}\frac{r_H}{r}. 
\label{potential}
\end{eqnarray}

The $1/r$ terms in $\Phi$ are expected from Newtonian gravity, and is
due to the boundary conditions for $\Phi$ at $r=r_H$; this is true for
the constant terms as well. The logarithmic term is due specifically
to the driving term $f(\mathbf{r})$, as expected. It is a 
long-range potential that extends out to $r_{II}$, and could potentially
explain the non-Newtonian interaction between galaxies and galactic
clusters in addition to the galactic rotation curves. The
$\rho_{II}^{1}(r) r^2$ term is due to the perturbation in the
background density, and contains terms $\sim 1/r^{1/2}$. It is due to both
the boundary terms in $\rho_{II}^{1}$ and the boundary conditions for
$\Phi(r)$. Finally, the $u^{2\alpha_\Lambda/(1+\alpha_\Lambda)}$ term in
Eq.~$(\ref{potential})$ is due to the background density 
$\rho_{\hbox{\scriptsize{asymp}}}$, with origins rooted in the rest mass
of the test particle. Note also that this term is less than $c^2$,
verifying that the nonrelativistic limit still holds.

The $c^2$ term in $\Phi(r)$ increases as $r^{6/5}$ for $\alpha_\Lambda =
3/2$, and would dominate the motion of test particles in the galaxy if
the extended GEOM depended on $\Phi(\mathbf{x})$ instead of
$\mathfrak{V}(\mathbf{x})$. Instead, it and the $r^2\rho_{II}^1$ term
are canceled by the additional density-dependent terms 
in Eq.~$(\ref{effPot})$. To see this, in Regions I and II
$\rho >> \Lambda_{DE}/2\pi$, and expanding Eq.~$(\ref{effPot})$ gives
\begin{equation}
\mathfrak{V}(\mathbf{x}) = \frac{1}{2}v_H^2
\left(\frac{r}{r_H}\right)^2-\frac{c^2}{2}+ 
\frac{c^2\chi}{2\alpha_\Lambda}\left(\frac{\Lambda_{DE}}{8\pi\rho_H}\right)^{\alpha_\Lambda}, 
\end{equation}
in Region I, while in Region II,
\begin{equation}
\mathfrak{V}(\mathbf{x}) = \Phi(\mathbf{x}) -\frac{\chi
  c^2(1+\alpha_\Lambda)^2}{4\alpha_\Lambda(1+3\alpha_\Lambda)}
\left(\frac{8\pi u^2}{\Lambda_{DE}}\right)
\left(\rho_{\hbox{\scriptsize{asymp}}} -
\alpha_\Lambda \rho^{1}_{II}\right),
\label{eff-pot}
\end{equation}
where we used $\rho_{II}^{1}/\rho_{\hbox{\scriptsize{asymp}}}<<1$ and
Eq.~$(\ref{amp})$. The last two terms of
Eq.~$(\ref{eff-pot})$ cancels the first two terms of
Eq.~$(\ref{potential})$, and $\mathfrak{V}(r)$ is
simply the remaining five terms in Eq.~$(\ref{potential})$;
the effective potential thus increases only logarithmically as $r$
increases. This is what is expected for a potential that determines
the motion of the stars in galactic rotation curves.

That the motion of stars in the galaxy is determined by
$\mathfrak{V}(\mathbf{x})$ and \textit{not} $\Phi(\mathbf{x})$ has far
reaching implications. The $c^2$ term in $\Phi(\mathbf{r})$ comes
from the background density $\rho_{\hbox{\scriptsize{asymp}}}$, and
thus the majority of the mass in a stationary galaxy \emph{does not
  contribute to the motion of test particles in the galaxy}. Rather,
it is the near-core density $\rho_{II}^{1}$ that contributes to
$\mathfrak{V}(\mathbf{x})$, which results in the very long-range,
logarithmic potential that is observed. As such, observations of the
rotational velocity curve of a galaxy will be able to determine the
perturbation on the background density, $\rho_{II}^{1}$, but not
$\rho_{\hbox{\scriptsize{asymp}}}$ itself. Consequently, since
$\rho_{\hbox{\scriptsize{asymp}}}(r)\gg \rho_{II}^{1}(r)$ when $r\gg
r_H$, \textit{the majority of the mass in the universe cannot be seen
  with these methods}. In particular, the motion of stars in galaxies
can only be used to estimate $\rho(r) -\rho_{\hbox{\scriptsize{asymp}}}(r)$;
the matter in $\rho_{\hbox{\scriptsize{asymp}}}(r)$ is present, but
cannot be ``seen'' in this way. 

This behavior is expected for a background density. In the
traditional theory of structure formation, the perturbation is off the
average density of matter of the universe. This density is usually
taken to be a constant, and thus cannot affect the motion of stars
within the galaxy. It also fits well with our interpretation that the 
extension of the GEOM is due to the replacement of the constant rest
mass $m$ with a curvature-dependent rest mass. While the rest
mass contributes to the Newtonian gravitational potential energy for
geodesic motion, it is a constant, and does not contribute to the dynamics of
the particle. We see a similar effect here. Our
$\rho_{\hbox{\scriptsize asymp}}$ is not a constant, however. It
\textit{increases} as $r\to r_H$, and this is a feature expected of cold
dark matter. Indeed, we will find below that $R_{200}$ is determined
primarily by $\rho_{\hbox{\scriptsize{asymp}}}$. As such, in
  the absence of all other forces the majority of the mass outside of
  the galaxy cannot be observed through its dynamics. Other means
would have to be used. 

\section{Cosmology}

In the previous sections, we have focused on analysing the structure of
a single galaxy. In this section, we will extrapolation these results
to the cosmological scale, and perform a cosmological check of our
theory. That this extrapolation can be done is based on the following
two observations 

First, recent measurements from WMAP and the Supernova Legacy Survey
put $\Omega_K = -0.011_{\pm 0.012}$, while WMAP and the HST key
project set $\Omega_K = -0.014_{\pm0.017}$ \cite{WMAP}. In both cases,
measurements have shown that the universe is essentially flat, and
WMAP's determination of $h=0.732_{-0.032}^{+0.031}$ was made under
this assumption, as is age of the universe $t_0$ at $13.73_{-0.15}^{+0.16}$
Gyr. As such, the largest distance between galaxies is $ct_0\equiv
\mathfrak{K}(\Omega) \lambda_{H}$, where $\mathfrak{K}(\Omega)$
depends on the details of how the universe evolves, and thus on its
thermal history; from measurements of both $h$ and $t_0$,
$\mathfrak{K}(\Omega)=1.03_{\pm0.05}$. 

Second, the density of matter of our model galaxy dies off
exponentially fast at $r_{II}$. The extent of the mass of matter in
the galaxy is thus fundamentally limited to $2r_{II}$. This size does
not depend on the detailed structure of the galaxy near its core; it
is inherent to the theory. Moreover, as we can express
\begin{equation}
r_{II} =
\left[\frac{8\pi\chi}{3\Omega_\Lambda(1+4^{1+\alpha_\Lambda})}\right]^{1/2}
\lambda_H 
\label{r-II-Lambda}
\end{equation}
where $\Omega_\Lambda = 0.716_{\pm 0.055}$ \cite{WMAP} is the
fractional density of Dark Energy in the universe, we
find that for $\alpha_\Lambda = 3/2$, $r_{II}=0.52\>\lambda_H$. Thus,
although the value of $\alpha_\Lambda$ was  
set to $3/2$ by the structure of the galaxy on a galactic scale, the
density distribution for the galaxy naturally cuts off at a radius equal
to half to Hubble scale, which is precisely what is expected from
cosmology.  

To accomplish the extrapolation, we use the properties of a
representative galaxy for the observed universe to construct our model
galaxy. This representative
galaxy could, in principal, be found by sectioning the observed
universe into three-dimensional, non-overlapping 
cells centered on a galaxy; given the spatial inhomogeneity of the
distribution of galaxies, these cells will not all be the same
size. Through a survey of these galactic cells, a representative
galaxy, with some average rotational velocity $v_H^{*}$ and core
radius $r^{*}_H$, can be found, and used to set the parameters for our
model galaxy. While such a survey has not yet been done, there
exists in the literature a large repository of measurements of
galactic rotation curves and core radii \cite{Blok-1,Cour,
  Math}. Taken as a whole, these 1393 galaxies are reasonably random,
and are likely representative of the observed universe at large. 

While we were able to estimate of $\alpha_\Lambda=3/2$ by looking at the galactic
structure, the accuracy of this estimate is unknown; comparison
with experiment will thus not possible. We instead \textit{require} that
$r_{II} = \mathfrak{K}(\Omega)\lambda_H/2$, which gives 
$\alpha_\Lambda$ as the solution of
$\mathfrak{K}(\Omega)^2(1+4^{1+\alpha_\Lambda}) =  
32\pi \chi(\alpha_\Lambda)/3\Omega_\Lambda$; this sets
$\alpha_\Lambda = 1.51_{\pm 0.11}$.

\subsection{$\sigma_8$, $R_{200}$, and Galactic Rotation Curves}

Linear fluctuations in the density are defined as
$\delta(\mathbf{r}) = \delta\rho/\langle\rho\rangle$
\cite{TurnerBook}, where 
$\langle\rho\rangle$ is the spatially-averaged density of matter in a
set volume. The rms fluctuation of $\delta$ is measured through
$\sigma_8^2 \equiv \langle (\delta(\mathbf{x})^2\rangle_8$, where the
subscript denotes a spatial average over $D_8$, a sphere 
with radius $8 h^{-1}$ Mpc. In calculating $\sigma_8$, we choose as our
origin the center of the unit cell containing the 
representative galaxy mentioned above. One estimate of the size of
a typical unit cell as $5 h^{-1}$ Mpc, the characteristic length scale
for the galaxy-galaxy correlation function \cite{TurnerBook}; as this is only
slightly smaller the $8 h^{-1}$, it is reasonable to
consider the density from only a single galaxy within $D_8$.

In our theory, $\rho(r)$ varies significantly across $D_8$. We thus
begin by calculating 
\begin{eqnarray}
\langle \rho\rangle_8 &\equiv& 
\langle \rho_H\theta(r_H-r)\rangle_8 +
\langle \rho_{\hbox{\scriptsize{asymp}}}(r)\theta(r-r_H)\rangle_8 + \langle
\rho_{II}^{1}(r)\theta(r-r_H)\rangle_8,
\nonumber
\\
&=&
\frac{3\Lambda_{DE}}{8\pi}\left(\frac{1+\alpha_\Lambda}{1+3\alpha_\Lambda}\right)
\Sigma(\alpha_\Lambda) {u}_8^{-2/(1+\alpha_\Lambda)}D(u_8)
\label{ave-rho}
\end{eqnarray}
where $\theta(x)$ is the step function, ${u}_8 = 8
h^{-1}\hbox{Mpc}/\chi^{1/2}\lambda_{DE}$, and 
\begin{equation}
\zeta=\frac{
2{u}_8^{-2\alpha_\Lambda/(1+\alpha_\Lambda)}
}{
\Sigma(\alpha_\Lambda)\chi(\alpha_\Lambda)
}
\left(\frac{v_H^{*}}{c}
\right)^2. 
\end{equation} 
The function 
\begin{eqnarray}
D(u_8) &\equiv & 1 - y_8^{\frac{1+3\alpha_\Lambda}{1+\alpha_\Lambda}} + 
\left[\frac{1+3\alpha_\Lambda}{1+\alpha_\Lambda}\right]\Bigg\{
\beta + \frac{3y_8(1+\alpha_\Lambda)^2}{2(1+3\alpha_\Lambda)}
\left[\nu_0\widetilde{C}_{\sin}-\frac{1}{2\nu_0}\widetilde{C}_{\cos}\right] -
\nonumber
\\
&{}&
\frac{3y_8^{\frac{1}{2}}(1+\alpha_\Lambda)^2}{2(1+3\alpha_\Lambda)}\left(
\left[\nu_0\widetilde{C}_{\cos}+\frac{1}{2}\widetilde{C}_{\sin}\right]
\sin\left[\nu_0\log y_8\right]+ \left[\nu_0\widetilde{C}_{\sin}-
  \frac{1}{2}\widetilde{C}_{\cos}\right]\cos\left[\nu_0\log y_8\right] 
\right)\Bigg\}, 
\nonumber
\\
\end{eqnarray}
where $y_8=u_H/u_8$, and
\begin{equation}
\widetilde{C}_{\cos} \equiv \frac{2}{3}\zeta
-\frac{1}{3}y^{2\alpha_\Lambda/(1+\alpha_\Lambda)}, \qquad
\nu_0\widetilde{C}_{\sin} \equiv \frac{7}{3}\zeta
-\frac{1}{6}\left(\frac{1+5\alpha_\Lambda}{1+\alpha_\Lambda}\right)
y^{2\alpha_\Lambda/(1+\alpha_\Lambda)}.
\end{equation}
Information on the structure of the galaxy is contained in $\zeta$.
As $\zeta\sim  5\times 10^{-3}$ for $v_H^{*}=200$ km/s, $D(u_8) \approx
1$, and it is the background density
$\rho_{\hbox{\scriptsize{asymp}}}$ that dominates
$\langle\rho\rangle_8$, and not the detail structure of the galaxy.  

This is not the case for $\langle\delta(\mathbf{x})^2\rangle_8$, which
involves the integration of $(\rho_{II})^2$ over $D_8$. Because
$\left(\rho_{II}^{1}\right)^2\sim 1/r^4$, it is now the behavior of
the density near the core that is relevant. Indeed, we find
\begin{eqnarray}
\sigma_8^2 =&{}& -1+
\frac{1}{3D(u_8)^2}\left(\frac{1+3\alpha_\Lambda}{1+\alpha_\Lambda}\right)^2
\Bigg\{\frac{1+\alpha_\Lambda}{3\alpha_\Lambda-1}
 +2\zeta\left(\frac{1+\alpha_\Lambda}{\alpha_\Lambda-1}\right)
\left(1-y_8^{(\alpha_\Lambda-1)/(\alpha_\Lambda+1)}\right)
+
\nonumber
\\
&{}&
\frac{3}{2}(1+\alpha_\Lambda)\left[\nu_0\widetilde{C}_{\sin}-
  \frac{(\alpha_\Lambda-3)}{2(1+\alpha_\Lambda)}
  \widetilde{C}_{\cos}\right]y_8^{(\alpha_\Lambda-1)/(\alpha_\Lambda+1)} -
\nonumber
\\
&{}&
\frac{3}{2}(1+\alpha_\Lambda)y_8^{1/2}\Bigg(
\left[\nu_0\widetilde{C}_{\cos}+
  \frac{(\alpha_\Lambda-3)}{2(1+\alpha_\Lambda)}
  \widetilde{C}_{\sin}\right]\sin[\nu_0\log y_8]+ 
\nonumber
\\
&{}&
\qquad\qquad\quad\>\>\qquad\left[\nu_0\widetilde{C}_{\sin}-
  \frac{(\alpha_\Lambda-3)}{2(1+\alpha_\Lambda)}
  \widetilde{C}_{\cos}\right]\cos[\nu_0\log y_8]\Bigg)+
\nonumber
\\
&{}&
\frac{4\zeta^2}{y_8}+\frac{3}{1+(1+3\alpha_\Lambda)/(1+\alpha_\Lambda)^2}
\left(\nu_0\widetilde{C}_{\sin}+
\frac{3}{2}\widetilde{C}_{\cos}\right)\frac{\zeta}{y_8} + 
\nonumber
\\
&{}&
\frac{9}{4}\left[\widetilde{C}_{\sin}^2+
\widetilde{C}_{\cos}^2+  
\frac{1}{1+\nu_0^2}\left(\widetilde{C}_{\sin}^2-
\widetilde{C}_{\cos}^2+2\nu_0 \widetilde{C}_{\cos}\widetilde{C}_{\sin}\right)\right]\frac{1}{y_8} 
\Bigg\},
\label{full-sig}
\\
\approx &{}&
\frac{4}{3}\frac{1}{(\alpha_\Lambda+1)(3\alpha_\Lambda-1)}+
\frac{1}{3}
\left(\frac{1+3\alpha_\Lambda}{1+\alpha_\Lambda}\right)^2\Bigg\{
\frac{8\alpha_\Lambda\zeta}{(\alpha_\Lambda-1)(3\alpha_\Lambda-1)}+
\frac{\zeta^2}{y_8}\Bigg[5+\frac{81}{4(1+\nu_0^2)}
\nonumber
\\
&{}&
\frac{10}{1+(1+3\alpha_\Lambda)/(1+\alpha_\Lambda)^2}\Bigg]\Bigg\},
\label{sigma-8}
\end{eqnarray}
where for Eq.~$(\ref{sigma-8})$ we have kept the lowest order terms in
$\zeta$ and $y_8$. The first term in Eq.~$(\ref{sigma-8})$ is due to
the background density $\rho_{\hbox{\scriptsize{asymp}}}$. It depends only on
$\alpha_\Lambda$, and contributes a set amount of 0.141 to $\sigma_8^2$
irrespective of the structure of the galaxy. The last term is due
primarily to the $1/r^2$ term in $\rho_{II}^{1}$, and is due to the
rotation curves. This term contributes the largest amount to 
$\sigma_8^2$, and depends explicitly on details of 
the structure of the galaxy through $v_H^{*}$ and $r_H^{*}$. 

While there have been a many studies of galactic rotation curves in
the literature, our need is for both the rotational velocity and the
core radius of galaxies. This requires both a measurement of of the
velocity as a function of the distance from the center of the galaxy,
and a fit of the data to some model of the velocity curve. To our
knowledge, this analysis has been done in four places in the
literature \footnote{A recent study \cite{McGa2005} gives fits to MOND
  rotation curves, but does not list values for $r_H$.}. While
each of the data sets were obtained with similar physical techniques,
there are distinct differences in their selection of galaxies, in the
exact experimental techniques used, and in their fits to rotation
curves (see \cite{Gent-2004} for a new method of deriving the rotation
  curves from H1 data). In fact, the Hubble constant used by each are
often different from one another, and from the value of 73.2 km/s/Mpc
given by WMAP. The reader is referred to the specific papers for
details on how these observations were made. Here, we only note the
following: 

\vskip 0.25in

\noindent \textit{de Blok et.~al.~Data Set:} 
  de Blok and coworkers made detailed measurements of 60 LSB galaxies
  \cite{McGa}, and fits to $v^{\hbox{\scriptsize{p-iso}}}(r)$ were
  done for 30 of them \cite{Blok-1}. Later, another set of
  measurements of 26 LSB galaxies 
  were made by de Blok and Bosma \cite{Blok-2}, of which 24 are
  different from the 30 listed in \cite{Blok-1}. Both the data for the
  30 original galaxies, and the 24 subsequent galaxies are
  used here. Although the authors used various models
  for determining the mass-to-light ratio in their measurements, we
  will use the data the comes from the minimum disk model, as this was
  the one model used for all of the galaxies in the set.
  
  While fits to the NFW density profile were made in
  \cite{Blok-1,Blok-2}, we are primarily concerned with the
  fits by the authors to the pseudoisothermal profile velocity curve
  Eq.~$(\ref{iso-curve})$. As de Blok and coworkers were chiefly concern
  was with the density parameter $\rho_H$ for the profile and $R_C$,
  the fits were made with these two parameters. Standard errors for 
  both $\rho_H$ and $R_C$ were calculated and given. Our concern is
  with the asymptotic value of $v_H$, however, and as this value is
  given by $\sqrt{4\pi G\rho_H r_C^2}$, we have calculated $v_H$ and
  its standard error from the published values of $\rho_H$ and $R_C$
  in \cite{Blok-1} to determine $v_H^{*}$ for this set. The authors
  used a value of 75 km/s/Mpc for the Hubble constant.  

\noindent\textit{CF Data Set:} In \cite{Cour}, Courteau presented
observations of the velocity curves for over 300 northern Sb-Sc UGC
galaxies, and determined $r_H$ for each by fitting the curves to three
different velocity curves, one of which is similar to the velocity
curve for the pseudoisothermal profile used by de Blok and coworkers, 
\begin{equation}
v^{\hbox{\scriptsize vcA}}(r)=\frac{2}{\pi}v_C \arctan\left(\frac{r}{r_t}\right).
\label{vcA}
\end{equation}
Like $v^{\hbox{\scriptsize p-iso}}(r)$, $v^{\hbox{\scriptsize
    vcA}}(r)$ can be approximated by the idealized velocity curve
$v^{\hbox{\scriptsize ideal}}(r)$ used in our analysis. In the limit
$r\gg r_t$, $v^{\hbox{\scriptsize vcA}}\approx v_C$, which sets $v_c =
v_H$. In the limit $r\ll r_t$, $v^{\hbox{\scriptsize vcA}}\approx
v_C(2r/\pi r_t)$, which sets $r_t = 2r_H/\pi$. Although there are
differences between $v^{\hbox{\scriptsize vcA}}(r)$ and
$v^{\hbox{\scriptsize p-iso}}(r)$ within the two limits, 
our analysis here is based on the idealized velocity curve and all
that is needed is the relationship between $r_H$, and $r_t$ or $R_C$.

The second fit was to a velocity curve where the steepness of
the transition from the hub and the asymptotic velocity curves could
be taken into account as well. Because of the work by de Blok and
coworkers in \cite{Blok-1}, our focus here will be on the velocity
curve Eq.~$(\ref{vcA})$, and not this curve. The second
velocity curve has a limiting form in the $r\to0$ limit that only
agrees with our idealized profile in one special case, and this case
does not hold for all the galaxies analyzed by Courteau using this profile. 

The third fit was to Persic, Salucci, and Stel's Universal Rotation Curve (URC)
\cite{Pers-1996}.  While the URC asymptotically approaches a constant
velocity, at small $r$ the URC has a $r^{0.66}$ behavior, which is
different from $v^{\hbox{\scriptsize{ideal}}}$, 
$v^{\hbox{\scriptsize{p-iso}}}$, and $v^{\hbox{\scriptsize{vcA}}}$,
all of which varies linearly in the small $r$ limit. We therefore
did not focus on this velocity curve here.

Values for $v_C$ and $r_t$ for 351 galaxies was obtained through the
VizieR service (http://vizier.u-strasbg.fr/viz-bin/VizieR). The great
majority of the rotation curves were based on single observations of
the galaxy; only 75 of these galaxies were measured multiple times,
with the majority of these galaxies observed twice. The data set
reposited at VizieR contained these multiple measurements. We have
averaged the value of $v_C$ and $r_t$ for the galaxy where there are
multiple measurements of the same galaxy. The standard error in the
repeated measurements of a single galaxy can be extremely large; this
was recognized in \cite{Cour}. A value of 70 km/s/Mpc was used for the
Hubble constant by the author.  

\noindent\textit{Mathewson et.~al.~Data Set:} In \cite{Math}, a survey
of the velocity curves of 1355 spiral galaxies in the southern sky was
performed. Later, the rotation curves for these observations were
derived in \cite{Pers-1995} after folding, deprojecting, and
smoothing the Mathewson data. Each of these velocity curves are due to
a single observation. Courteau performed a fit of Mathewson's
observations to the velocity curve Eq.~$(\ref{vcA})$ for 958 of the
galaxies in \cite{Cour} using a Hubble constant of $70$ km/s/Mpc. The
results of this analysis is reposited in VizieR as well. 

\noindent\textit{Rubin et.~al.~Data Set:}  In the early 1980s, Rubin and
  coworkers \cite{Rubin1980, Rubin1982, Burs, Rubin1985} presented a
  detailed study of the rotation curves of 16 Sa, 23 Sb, and 21 Sc
  galaxies. This was \textit{not} a random sampling of Sa, Sb, and Sc
  galaxies. Rather, these galaxies were
  deliberately chosen to span a specified range of Sa, Sb, and Sc
  galaxies, and as stated in \cite{Burs}, averaging values of properties of
  galaxies in this data set would have little meaning. These measurements
  can contribute to the combined data set of all four measurements,
  however, which is why we have included them in our analysis.
  While values for the core radius were not given, measurements of the
  rotational velocity as a function of the distance to the center of
  the galaxy were; we are able to fit the data to the same
  pseudoisothermal rotation curve Eq.~$(\ref{iso-curve})$ used by de
  Blok, et.~al. Results of this fit is given in \textbf{Appendix
    A.1}. A Hubble constant of 50 km/s/Mpc was used by the authors. 

\vskip12 pt

Wanting to be as unbiased and as inclusive as possible, we have deliberately
\textit{not} culled through the data sets to select the cleanest of the
rotation curves. Nevertheless, we have had to removed the data for 27
galaxies from our analysis. A list of these galaxies and the reason
why they were removed are given in \textbf{Appendix B}, where we have
listed other peculiarities found with the data sets as well. 

The $v_H^{*}$ and $r_H^{*}$, and standard error for each, were calculated
for each of the four data sets considered here. While $v_H$ is easily
identified for all four, determination of $r_H$ is more complicated. For the
de Blok et.~al.~data set, published values of $R_C$ was first scaled
by $75/73.2$ to account for differences in the Hubble constant used by
the authors, and the current value of $73.2$ km/s/Mpc measured by
\cite{WMAP}. Then $r_H$ is obtained by using $r_H=\sqrt{3}R_C$. A
similar calculation was made using the calculated values of $R_C$ from
\textbf{Appendix A.1}, but using $50/73.2$ instead of $75/73.2$ to
account for differences in Hubble constants. For the CF and Mathewson
et.~al.~data sets, published values of $r_t$ are first scaled by
$70/73.2$ to account for differences in Hubble constants, and 
$r_H$ is now obtained using $r_H=\pi r_t/2$. A fifth data set is then
constructed by combining the data from these four data sets. For each
data set, $v^{*}_H$ and $r^{*}_H$  are then used in
Eq.~$(\ref{full-sig})$ to calculate $\sigma_8$; numerical 
derivatives of $\sigma_8$ were then used to calculate its standard
error. Not surprisingly, $\Delta\sigma_8$ is dominated by the standard
error in $\alpha_\Lambda$. 

Results of these calculations are giving in Table \ref{summary}, along
with the t-test comparison of the calculated $\sigma_8$ and
$\Delta\sigma_8$ with the value $0.761_{-0.048}^{+0.049}$ from
\cite{WMAP}. Surprising, four of the five data sets give a value for
$\sigma_8$ that is within two-sigma of the WMAP value; they thus agree
with the WMAP value at the 95\% confidence level. The only data  
set that differs significantly from the WMAP value is the Rubin
et.~al.~set, and it is known that for this set the selected galaxies
are not representative of Sa, Sb. and Sc galaxies; this disagreement
is thus not surprising.

\begin{table}
{\centering
\begin{tabular}{l|rrrr|rrrr}
\hline
\textit{Data Set} &   \hskip0.1in $v^{*}_H$ &   \hskip0.1in $\Delta v^{*}_H$ &   \hskip0.1in $r^{*}_H$ &   \hskip0.1in $\Delta r^{*}_H$
&   $\qquad\sigma_8$ &  $\qquad\Delta \sigma_8$ &   \hskip0.1in \textit{t-test} &   \\
\hline
deBlok et.~al. (53)           &  119.0 &  6.8 &   3.62 &  0.33 & 0.613 &  0.097 &  1.36 \\
CF (348)                      &  179.1 &  2.9 &   7.43 &  0.35 & 0.84 &  0.18 &  0.43 \\
Mathewson et.~al.~(935)       &  169.5 &  1.9 &  15.19 &  0.42 & 0.625 &  0.089 &  1.34 \\
Rubin et.~al. (57)            &  223.3 &  7.6 &   1.24 &  0.14 & 2.79  &  0.82  &  2.46 \\
Combined (1393)               &  172.1 &  1.6 &  11.82 &  0.30 & 0.68 &  0.11 &  0.70 \\
\hline
\end{tabular}
\par}
\caption{The $v_H^{*}$ (km/s), $r_H^{*}$ (kps), and resultant
  $\sigma_8$, $\Delta\sigma_8$, and t-test comparison with the WMAP
  value of $\sigma_8$. The number of data points in each data set is
  listed in the parentheses.}  
\label{summary}
\end{table}  

While the URC has a different power-law behavior at small $r$ than
$v^{\hbox{\scriptsize{ideal}}}$, $v^{\hbox{\scriptsize{p-iso}}}$, or
$v^{\hbox{\scriptsize{vcA}}}$, the difference is small enough that it
is unknown how $\sigma_8$ will change if the URC is used in its
calculation instead of the $v^{\hbox{\scriptsize{ideal}}}$ used
here. We leave this for future research.

Given $v_H^{*}$ and $r_H^{*}$, it is possible to calculate
$R_{200}$ by setting $u_8\to u_{200}=R_{200}/\chi^{1/2}\lambda_{DE}$ in
Eq.~$(\ref{ave-rho})$, and using $v_H^{*}= 172.1$ km/s and $r_H^{*}=11.82$
kpc from the Combine data set. We numerically solved for this
radius and found that $R_{200} =  270_{\pm130}$ 
kpc, with the large spread coming primarily from the uncertainty in
$\alpha_\Lambda$. 

\subsection{Estimating the Fractional Densities of Matter}

Because $\rho_{\hbox{\scriptsize{asymp}}}$ is an asymptotic solution and has
the same form irrespective the the detail shape of the galaxies, we
can estimate $\Omega_{\hbox{\scriptsize asymp}}$ by averaging
$\rho_{\hbox{\scriptsize{asymp}}}(r)$ over a sphere of radius
$\lambda_H/2=r_{II}$,
\begin{equation}
\Omega_{\hbox{\scriptsize{asymp}}}\equiv \frac{\langle
  \rho_{\hbox{\scriptsize{asymp}}}\rangle_{r_{II}}}{\rho_c} \approx  
\frac{3\Omega_\Lambda}{8\pi} \left(\frac{1+\alpha_\Lambda}{1+3\alpha_\Lambda}\right)
\left[2
\frac{(1+3\alpha_\Lambda)}{(1+\alpha_\Lambda)^2}\left(1+4^{1+\alpha_\Lambda}\right)\right]^{1/(1+\alpha_\Lambda)}
+ \mathcal{O}\left(\frac{r_H}{\lambda_H}\right)^3, 
\label{ratio}
\end{equation}
where $\rho_c$ is the critical density of the universe, and we have
used Eq.~$(\ref{r-II-Lambda})$. Thus, the ratio
$\Omega_{\hbox{\scriptsize asymp}}/\Omega_\Lambda$ depends only on
the dimensionality and symmetry of spacetime, and the exponent
$\alpha_\Lambda$. Numerically, we find $\Omega_{\hbox{\scriptsize{asymp}}} =
0.197_{\pm0.017}$. 

In performing this average we have implicitly
assumed that there is only a single galaxy within the sphere, which is
a gross under counting of the number of galaxies in the
universe. Note, however, that $\rho_{\hbox{\scriptsize asymp}}(r)$ is an
asymptotic solution, and $\rho_{II}^{1}(r)$ is a perturbation off
$\rho_{\hbox{\scriptsize asymp}}(r)$ that dies off
rapidly with distance. While additional galaxies within the
sphere may change the detail form of $\rho_{\hbox{\scriptsize asymp}}$, these
changes are expected to be equally short ranged; we thus expect
Eq.~$(\ref{ratio})$ to be an adequate estimate of 
$\Omega_{\hbox{\scriptsize asymp}}$. 

Such is not the case for
$\Omega_{\hbox{\scriptsize{Dyn}}}$. Calculating
$\Omega_{\hbox{\scriptsize Dyn}}$ directly by averaging
$\rho-\rho_{\hbox{\scriptsize asymp}}$ would require knowing both 
the detail structure of galaxies, and the distribution of galaxies
within the sphere. Instead, we note that $\Omega_m =
\Omega_{\hbox{\scriptsize asymp}}+\Omega_{\hbox{\scriptsize Dyn}}$,
and using the value $\Omega_m=0.239_{-0.026}^{+0.025}$ from WMAP,
find that
$\Omega_{\hbox{\scriptsize{Dyn}}}=0.041^{+0.030}_{-0.031}$. Thus, only 
a small fraction of the matter in the universe can be seen through
their dynamics. 

\section{Concluding Remarks}

Given how sensitive of our expression for $\sigma_8$ is dependent on
$v_H^{*}$, $r_H^{*}$, and $\alpha_\Lambda$, that our predicted values of
$\sigma_8$ is within experimental error of its measured value is
surprising. This is especially true as the data used in calculating
$\sigma_8$ was taken by four different groups over 
a period of 25 years, and for purposes that have no connection
whatsoever with our analysis. Even in the absence of a direct
experimental search for $\alpha_\Lambda$, this provides a compelling
argument for the validity of our extension of the GEOM, and its impact
on structure formation. This agreement also supports our free energy
conjecture; the calculated $\sigma_8^2$ would be very different if
$\beta=3$, say, were used in calculating $\sigma_8^2$ instead of
$\beta=2$. 

Direct detection and measurement of $\alpha_\Lambda$ through terrestrial
experiments may be possible in the near future. As mentioned in
Sections \textbf{II.D} and \textbf{III.B}, at a value of $1.51$ the
exponent $\alpha_\Lambda$ is likely small enough that the effects of
the additional terms in the extended GEOM may soon be detectable. 

Interestingly, $\Omega_m-\Omega_{B} = 0.196_{-0.026}^{+0.025}$ is
nearly equal to $\Omega_{\hbox{\scriptsize
    asymp}}$ in value. Correspondingly, $\Omega_B$  
\cite{WMAP} is nearly equal to $\Omega_{\hbox{\scriptsize
    Dyn}}$. It would be tempting to identify 
$\Omega_{\hbox{\scriptsize asymp}}$ with the fractional density of
nonbaryonic (dark) matter in the universe, especially since matter in
$\rho_{\hbox{\scriptsize{asymp}}}(r)$ does not participate in the particle
dynamics, and is not ``visible'' to measurements that inferred
mass through particle motion under gravity. That
$\Omega_{\hbox{\scriptsize Dyn}}$ would then be identified with
$\Omega_B$ is consistent with the observation 
that most of the mass that has been inferred through gravitational
dynamics indeed consists of baryons. We did not differentiate between
normal and dark matter in our theory, however. Without a specific
mechanism for funneling nonbaryonic matter into
$\rho_{\hbox{\scriptsize{asymp}}}$ and baryonic matter into
$\rho-\rho_{\hbox{\scriptsize{asymp}}}$, we 
cannot at this point rule out the possibility that
$\Omega_m-\Omega_B=\Omega_{\hbox{\scriptsize{asymp}}}$ and
$\Omega_B\approx\Omega_{\hbox{\scriptsize Dyn}}$ is a numerical
accident. 

\appendix{}

\section{Fits to Data}

In \cite{Rubin1985}, measurements of the rotational velocity as a
function of radius for 60 Sa, Sb and Sc spiral galaxies are given,
allowing a fit of this data to $v^{p-iso}(r)$. However,
instead of fitting to $v^{\hbox{\scriptsize p-iso}}(r)$ directly as
is done in \cite{Blok-1}, it is more convenient to fit the data to
$(v^{\hbox{\scriptsize p-iso}}(r))^2$. As we are interested in the
asymptotic velocity $v_H$ instead of the density 
parameter for the pseudoisothermal profile, ours is a two
parameter, $(v_H,R_C)$, least-squares fit to $(v^{\hbox{\scriptsize
    p-iso}}(r))^2 = v_H^2 c(r)$, where  
\begin{equation}
c(r) = 1-\frac{R_C}{r} \arctan\left(\frac{r}{R_C}\right).
\end{equation}
It uses the variance
\begin{equation}
\sigma_{(v^{\hbox{\scriptsize p-iso}})^2}^2 \equiv
\frac{1}{N-2}\sum_{n=1}^N [(v^{\hbox{\scriptsize p-iso}}_n)^2 - v_H^2c(r_n)]^2,
\label{D}
\end{equation}
where $\{(v_n^{\hbox{\scriptsize{p-iso}}}, r_n)\}$ is the set of
velocity verses radius measurements for a galaxy with a total number
of data points $N$. Minimizing with respect to $v_H^2$ gives
\begin{equation}
v_H^2 = \frac{\langle (v^{\hbox{\scriptsize p-iso}}_n)^2 c(r_n)\rangle}{\langle
  c(r_n)^2\rangle},
\label{v-infty}
\end{equation}
where $\langle \cdots\rangle$ denotes an average over the data
points. Minimization with respect to $R_C$ gives the implicit
equation 
\begin{equation}
0=\langle (v^{\hbox{\scriptsize p-iso}}_n)^2 c(r_n)\rangle 
\left\langle\frac{c(r_n) r_n^2}{r_n^2+R_C^2}\right\rangle
-\langle c(r_n)^2\rangle 
\left\langle\frac{(v^{\hbox{\scriptsize p-iso}}_n)^2
  r_n^2}{r_n^2+R_C^2}\right\rangle. 
\label{R-C}
\end{equation}
Instead of solving Eq.~$(\ref{R-C})$ directly, we substitute
Eq.~$(\ref{v-infty})$ into Eq.~$(\ref{D})$, and find iteratively the
$R_C$ that minimizes $\sigma^2_{(v^{\hbox{\scriptsize p-iso}})^2}$. The
value for $v_H^2$ is then found through Eq.~$(\ref{v-infty})$.

Standard errors for $v_H^2$ and $R_C$ can be
found directly. Taking the implicit
derivative of Eq.~$(\ref{R-C})$, 
\begin{equation}
\frac{\partial R_C}{\partial (v^{\hbox{\scriptsize p-iso}}_i)^2} =
\frac{R_C}{N\Delta} 
\left\{
    \frac{\langle c(r_n)^2\rangle r_i^2}{r_i^2+R_C^2} -c(r_i)\left\langle\frac{c(r_n)r_n^2}{r_n^2+R_C^2}\right\rangle
\right\},
\end{equation}
where
\begin{eqnarray}
\Delta \equiv &{}& 
2\langle (v^{\hbox{\scriptsize p-iso}}_n)^2c(r_n)\rangle
\left\langle\frac{c(r_n)r_n^4}{(r_n^2+R_C^2)^2}\right\rangle -
2\langle c(r_n)^2\rangle
\left\langle\frac{(v^{\hbox{\scriptsize p-iso}}_n)^2
  r_n^4}{(r_n^2+R_C^2)^2}\right\rangle + 
\nonumber
\\
&{}&
\left\langle \frac{c(r_n)r_n^2}{r_n^2+R_C^2}\right\rangle
\left\langle\frac{(v^{\hbox{\scriptsize p-iso}}_n)^2
  r_n^2}{r_n^2+R_C^2}\right\rangle - 
\langle (v^{\hbox{\scriptsize p-iso}}_n)^2 c(r_n)\rangle
\left\langle\frac{r_n^4}{(r_n^2+R_C^2)^2}\right\rangle.
\end{eqnarray}
The standard error $\sigma_{R_C}$ in $R_C$ is then
\begin{equation}
\sigma_{R_C}= \frac{R_C\> \sigma_{(v^{\hbox{\scriptsize p-iso}}_n)^2}\sqrt{\langle
    c(r_n)^2\rangle}}{\Delta\sqrt{N}} \left\{\langle
c(r_n)^2\rangle\left\langle\frac{r_n^4}{(r_n^2+R_C^2)^2}\right\rangle 
-\left\langle\frac{c(r_n)r_n^2}{r_n^2+R_C^2}\right\rangle^2 \right\}^{1/2}.
\end{equation}

For the standard error in $v_H^2$, we use Eq.~$(\ref{v-infty})$ and
find 
\begin{equation}
\frac{\partial v_H^2}{\partial (v^{\hbox{\scriptsize p-iso}}_i)^2} =
\frac{c(r_i)}{N\langle c(r_n)^2\rangle}
-\left(1 + \frac{1}{v_H^2\langle c(r_n)^2\rangle}
\left\langle\frac{v^{\hbox{\scriptsize{p-iso}}}_nf_n^2}{R_C^2+r_n^2}\right\rangle 
-
\langle\frac{2}{c(r_n)^2}\rangle
\left\langle\frac{c(r_n)r_n^2}{R_c^2+r_n^2}\right\rangle 
\right)\frac{v_H^2}{R_C} 
\frac{\partial R_C}{\partial (v^{\hbox{\scriptsize{p-iso}}})^2}, 
\end{equation}
resulting in a standard error in $v_H$ of
\begin{equation}
\sigma_{v_H}=\frac{1}{2}\left[\frac{\sigma^2_{(v^{\hbox{\scriptsize p-iso}}_n)^2}}{
  Nv_H^2\langle c(r_n)^2\rangle} + 
\left(1 + \frac{1}{v_H^2\langle c(r_n)^2\rangle}
\left\langle\frac{v^{\hbox{\scriptsize{p-iso}}}_nr_n^2}{R_C^2+r_n^2}\right\rangle 
-
\frac{2}{\langle c(r_n)^2\rangle}\left\langle\frac{c(r_n)r_n^2}{R_c^2+r_n^2}\right\rangle
\right)^2\frac{\sigma_{R_C}^2}{R_c^2}
\right]^{1/2}. 
\end{equation}

Our fits of the Rubin et.~al.~data are tabulated in Table
\ref{Rubin}. The base data from \cite{Rubin1985} was based on a
Hubble constant of $50$ km/s/Mpc, and the results given in Table
\ref{Rubin} are for this value of the Hubble constant. Of the 60
galaxies from \cite{Rubin1985}, NGC 6314 and IC 724 could 
not be fitted to a nonzero $R_C$, while the fit for NGC  2608 resulted
in a $R_C$ that is less than $0.01$. 

\begin{table}
{
\centering
\begin{tabular}{l|rrrr||cl|rrrr}
\hline
Galaxy &   \hskip0.175in  $R_C$ & \hskip0.175in$\Delta R_C$ & $\quad\qquad v_H$ &   \hskip0.175in $\Delta v_H$ & & Galaxy  &  \hskip0.175in$R_C$ &   \hskip0.175in $\Delta R_C$ & $\quad\qquad v_H$ & \hskip0.175in $\Delta v_H$   \\
\hline
NGC 1024 &   0.27 &   0.14 &   229.42 &    9.77 &    &   NGC 4800 &   0.18 &   0.06 &   171.56 &   3.42    \\
NGC 1357 &   0.52 &   0.14 &   268.19 &   16.27 &    &   NGC 7083 &   0.89 &   0.14 &   226.51 &   2.27    \\
NGC 2639 &   1.02 &   0.34 &   337.69 &   31.31 &    &   NGC 7171 &   2.25 &   0.36 &   251.35 &   6.47    \\
NGC 2775 &   0.40 &   0.17 &   298.98 &    8.90 &    &   NGC 7217 &   0.19 &   0.10 &   275.21 &   6.56    \\
NGC 2844 &   0.41 &   0.09 &   167.50 &   18.93 &    &   NGC 7537 &   0.80 &   0.10 &   150.06 &   2.35    \\
NGC 3281 &   0.44 &   0.05 &   211.32 &   26.51 &    &   NGC 7606 &   1.40 &   0.30 &   279.29 &   4.11    \\
NGC 3593 &   0.16 &   0.08 &   115.28 &   14.01 &    &   UGC 11810&   1.54 &   0.38 &   193.28 &   3.85    \\
NGC 3898 &   0.53 &   0.06 &   254.76 &   28.73 &    &   UGC 12810&   3.22 &   0.35 &   245.73 &   1.47    \\
NGC 4378 &   0.13 &   0.06 &   307.61 &   26.60 &    &   NGC 701  &   2.49 &   0.58 &   188.78 &   4.86    \\
NGC 4419 &   0.63 &   0.03 &   211.55 &    2.33 &    &   NGC 753  &   0.31 &   0.11 &   208.50 &   3.57    \\
NGC 4594 &   1.65 &   0.30 &   397.24 &   10.15 &    &   NGC 801  &   0.79 &   0.16 &   227.64 &   4.06    \\
NGC 4698 &   1.85 &   0.47 &   284.96 &    6.34 &    &   NGC 1035 &   1.24 &   0.09 &   150.62 &   1.26    \\
NGC 4845 &   0.11 &   0.07 &   187.54 &    0.07 &    &   NGC 1087 &   0.54 &   0.10 &   131.91 &   2.54    \\
UGC 10205&   2.19 &   0.27 &   272.34 &    4.07 &    &   NGC 1421 &   0.54 &   0.13 &   176.42 &   3.94    \\
NGC 1085 &   0.29 &   0.05 &   307.02 &    2.11 &    &   NGC 2715 &   1.10 &   0.22 &   151.47 &   2.93    \\
NGC 1325 &   1.80 &   0.28 &   195.55 &    2.67 &    &   NGC 2742 &   1.10 &   0.16 &   181.86 &   2.36    \\
NGC 1353 &   0.36 &   0.18 &   218.48 &    8.30 &    &   NGC 2998 &   1.08 &   0.22 &   213.85 &   3.22    \\
NGC 1417 &   0.40 &   0.05 &   278.87 &    2.36 &    &   NGC 3495 &   3.11 &   0.46 &   206.75 &   3.22    \\
NGC 1515 &   0.06 &   0.10 &   178.35 &   10.03 &    &   NGC 3672 &   1.74 &   0.24 &   208.11 &   4.03    \\
NGC 1620 &   1.73 &   0.25 &   241.62 &    3.14 &    &   NGC 4062 &   0.79 &   0.13 &   167.88 &   2.65    \\
NGC 2590 &   1.30 &   0.54 &   255.24 &    5.33 &    &   NGC 4321 &   0.79 &   0.35 &   208.24 &   5.42    \\
NGC 2708 &   1.91 &   0.68 &   269.92 &    9.45 &    &   NGC 4605 &   0.97 &   0.32 &   112.62 &   3.42    \\
NGC 2815 &   1.91 &   0.68 &   269.92 &    9.45 &    &   NGC 4682 &   1.17 &   0.23 &   181.17 &   2.97    \\
NGC 3054 &   2.41 &   0.56 &   259.10 &    8.30 &    &   NGC 7541 &   0.21 &   0.16 &   195.04 &   5.94    \\
NGC 3067 &   0.76 &   0.06 &   156.80 &    1.22 &    &   NGC 7664 &   0.65 &   0.14 &   196.05 &   3.07    \\
NGC 3145 &   0.15 &   0.07 &   257.00 &    4.84 &    &   IC 467   &   1.64 &   0.33 &   152.42 &   3.26    \\
NGC 3200 &   0.42 &   0.09 &   266.07 &    5.43 &    &   UGC 2885 &   0.06 &   0.10 &   266.22 &   5.88    \\
NGC 3223 &   1.35 &   0.23 &   275.29 &    5.51 &    &   UGC 3691 &   3.04 &   0.33 &   229.42 &   1.31    \\
NGC 4448 &   0.59 &   0.11 &   207.02 &    1.98 & & & & & & {} \\
\hline
\end{tabular}
\par
}
\caption{Fitted values of $R_C$ (kpc) and $v_H$ (km/s), and their errors for
  Rubin et.~al.~data.}  
\label{Rubin}
\end{table}

\section{Data sets}

For the de Blok et.~al.~data set, the galaxy F568-3 was analysed
twice; we use analysis of F568-3 given by the authors
in \cite{Blok-1}. In de Blok and Bosma \cite{Blok-2}, two of the
galaxies, F563-1 and U5750, also appeared in \cite{Blok-1}; we used
the values from \cite{Blok-1} for these galaxies in our analysis.
The radius for DDO185 from \cite{Blok-1} was not determined, and we
could not include this data point in our analysis. Thus, out of
56 possible galaxies, 53 were used.  

For the Rubin et.~al.~data set, we could not find a nonzero radius for
two galaxies, and one galaxy had a radius less than 0.01 kpc. As this
radius was smaller than the resolution of their observations, this
data point was not included. A total of 57 galaxies were thus used from
\cite{Rubin1985}.  

For the CF and Mathewson et.~al.~data sets, the vast majority of the data  
were based on single observations. We therefore had greater leeway in
cleaning up this data, but even here we were circumspect. First, 75
galaxies in \cite{Cour} were observed multiple times. Of these, the
galaxies UGC 7234 and UGC 10096 had listed an asymptotic velocity for
one of the observations that was opposite from the 
measured asymptotic velocity for the others. We assumed that this
was a typographical error, and the sign of the rotational velocity for
the anomalous velocity is reversed. Second, five galaxies in
the CF and Mathewson et.~al.~data sets had a $r_H=0$, one galaxy had a radius
core that was 11-sigma out, and three galaxies had a $v_H$
that exceeded 8,000 km/s. These are likely indications that the data was not
sufficiently accurate to allow for a fit of the velocity curve, and
they were removed. Finally, given that there are only 1393 galaxies
combined in the data sets, if a galaxy had a $v_H$ or a $r_H$ that was
six-sigma or more out from the mean, they were removed. In the end,
348 galaxies were used in the CF data set, and 935 galaxies were used
in the Mathewson et.~al.~data set. A summary of the data points not
used in our analysis is given in Table \ref{removed}. 

\begin{table}
{\centering
\begin{tabular}{c|c|c||c|c|c}
\hline
\textit{Data Set} & \textit{Data Removed} & \textit{Reason} &\textit{Data Set} & \textit{Data Removed} & \textit{Reason} \\
\hline
de Blok et.~al. & DDO185      &   $r_H=\infty$                            & {} &   ESO 243-G34 &   $v_H$ is 5 $\sigma$ out      \\   
Rubin et.~al.  & NGC 6314    &   $r_H = 0$                               & {} &   ESO 317-G41 &   $r_H = 0$               \\  
{}              & IC 724      &   $r_H = 0$                               & {} &   ESO 358-G9  &   $v_H$ is 6 $\sigma$ out \\  
{}              & NGC 2608    &   $r_H < 0.01$                            & {} &   ESO 435-G25 &   $v_H$ is 5 $\sigma$ out \\   
CF              & UGC 6534    &   $v_H$ is 35 $\sigma$ out                & {} &   ESO 467-G12 &   $r_H = 0$               \\  
{}              & UGC 12543   &   $v_H$ is 11 $\sigma$ out                & {} &   ESO 554-G28 &   $v_H$ is 6 $\sigma$ out \\  
Mathewson et.~al.~      & ESO 140-G28 &   $v_H > 8,000$ km/s              & {} &   ESO 60-G24  &   $v_H$ is 10 $\sigma$ out\\
{}              & ESO 481-G30 &   $v_H > 24,000$ km/s                     & {} &   ESO 359-G6  &   $r_H$ is 11 $\sigma$ out\\
{}              & ESO 443-G42 &   $v_H > 94,000$ km/s                     & {} &   ESO 481-G21 &   $r_H$ is 6 $\sigma$ out    \\
{}              & ESO 108-G19 &   $r_H = 0$                               & {} &   UGCA 394    &   $r_H$ is 7 $\sigma$ out \\
{}              & ESO 141-G34 &   $r_H = 0$                               & {} &   ESO 298-G15 &   $ r_H$ is 7 $\sigma$ out\\
{}              & ESO 21-G5   &   $r_H$ is 6 $\sigma$ out {}              & {} &   ESO 545-G3  &   $r_H$ is 7 $\sigma$ out \\
{}              & ESO 548-G21 &   $r_H$ is 7 $\sigma$ out {}              & {} &   ESO 404-G18 &   $r_H$ is 9 $\sigma$ out \\
{}              & NGC 7591    &   $r_H = 0$                               & {} &   {}          &   {}              \\
\hline
\end{tabular}
\par}
\centering
\caption{Listed are the galaxies removed from the data sets used in
  our analysis along with the reason for their removal.}
\label{removed}
\end{table}


\begin{acknowledgments}

The author would like to thank John Garrison for the numerous suggestions,
comments, and the support he has given of his time while this research
was being done. His efforts have helped guide it, and have elucidated
many of the arguments given here. The author would also like to thank
K.-W. Ng, H. T. Cho, and Clifford Richardson for their comments and
criticisms as this research was done.

\end{acknowledgments}

\end{document}